\documentclass[preprint2]{aastex}

\usepackage{color}
\definecolor{Dred}{rgb}{0.312,0.070,0.070}
\definecolor{Dblue}{rgb}{0.070,0.070,0.312}
\definecolor{Dgreen}{rgb}{0.070,0.312,0.070}
\definecolor{Db}{rgb}    {0.050,0.0,0.320}

\newcounter{note}
\let\oldmarginpar\marginpar
\renewcommand\marginpar[1]{\-\oldmarginpar[\raggedleft\footnotesize #1]{\raggedright\footnotesize #1}}

\shorttitle{iMOGABA I}
\shortauthors{Lee et al.}

\begin{document}

\title{Interferometric Monitoring of Gamma-ray Bright AGNs I:
Results of Single-epoch Multifrequency Observations}

\author{
Sang-Sung Lee\altaffilmark{1,2},       
Kiyoaki Wajima\altaffilmark{1},        %
Juan-Carlos Algaba\altaffilmark{1},    %
Guang-Yao Zhao\altaffilmark{1},        %
Jeffrey A. Hodgson\altaffilmark{1},        %
Dae-Won Kim\altaffilmark{3},           %
Jongho Park\altaffilmark{3},           %
Jae-Young Kim\altaffilmark{4},
Atsushi Miyazaki\altaffilmark{5},
Do-Young Byun\altaffilmark{1,2},       %
Sincheol Kang\altaffilmark{1,2},       %
Jeong-Sook Kim\altaffilmark{6},
Soon-Wook Kim\altaffilmark{1,2},
Motoki Kino\altaffilmark{1},
and Sascha Trippe\altaffilmark{3}
}

\altaffiltext{1}{Korea Astronomy and Space Science Institute, 
776 Daedeok-daero, Yuseong-gu, Daejeon 34055,
Republic of Korea; sslee@kasi.re.kr}

\altaffiltext{2}{Korea University of Science and Technology,
217 Gajeong-ro, Yuseong-gu, Daejeon 34113,
Republic of Korea}

\altaffiltext{3}{Department of Physics and Astronomy,
Seoul National University, 
1 Gwanak-ro, Gwanak-gu, Seoul 08826,
Republic of Korea}

\altaffiltext{4}{Max-Planck-Institut f\"ur Radioastronomie,
Auf dem H\"ugel 69, 53121 Bonn,
Germany}

\altaffiltext{5}{Japan Space Forum,
3-2-1, Kandasurugadai, Chiyoda-ku,
Tokyo 101-0062
Japan}

\altaffiltext{6}{National Astronomical Observatory of Japan,
2211 Osawa, Mitaka, Tokyo 1818588,
Japan}

\begin{abstract}
We present results of single-epoch very long baseline interferometry (VLBI)
observations of gamma-ray bright active galactic nuclei (AGNs) using the
Korean VLBI Network (KVN) at 22, 43, 86, and 129~GHz
bands, which are part of a KVN key science program,
Interferometric Monitoring of Gamma-ray Bright AGNs (iMOGABA). We selected a
total of 34 radio-loud AGNs of which 30 sources are
gamma-ray bright AGNs with flux densities of $>6\times10^{-10}$~ph~cm$^{-2}$~s$^{-1}$. Single-epoch
multi-frequency VLBI observations of the target sources were conducted during
a 24-hr session on 2013 November 19 and 20. All
observed sources were detected and imaged at all frequency bands
with or without a frequency phase transfer technique which enabled
the imaging of 12 faint sources at 129~GHz, except for one source.
Many of the target sources are resolved on
milliarcsecond scales, yielding a core-jet structure with the VLBI core
dominating the synchrotron emission on the milliarcsecond scale. CLEAN flux
densities of the target sources are 0.43-28~Jy, 0.32-21~Jy, 0.18-11~Jy, and
0.35-8.0~Jy in the 22, 43, 86, and 129~GHz bands, respectively.
Spectra of the target
sources become steeper at higher frequency with the mean of
the spectral indices of -0.40, -0.62, and -1.00 in the
22-43~GHz, 43-86~GHz and 86-129~GHz bands, respectively, implying that the target
sources become optically thin at higher frequency (e.g., 86-129GHz).
\end{abstract}


\section{Introduction}

Active galactic nuclei (AGNs) are the most powerful and violent
celestial objects in the Universe. 
AGNs emit a wide range of electromagnetic radiation
in the radio to gamma-ray range, producing huge luminosity ($\sim10^{42}-10^{48}$~erg~s$^{-1}$)
in very compact emitting regions on sub-pc scales; 
their emission is often highly variable and polarized~\citep[see e.g.,][]{kro99}.
The AGNs are believed to have supermassive black holes (SMBH) in the centers,
accretion disks around the SMBHs, 
and relativistic jets launched perpendicular to the accretion disk plane into interstellar space.
Relativistic jets of the radio-loud AGNs often show
variability in brightness (e.g., in radio, optical, etc.)
on timescales of hours to years,
and morphological changes on milliarcsecond (mas)
scale~\citep[see e.g., ][for review]{boe+12}.

Very rapid and strong variabilities (or flares) in the gamma-ray range
from radio-loud AGNs
have been observed during intensive gamma-ray monitoring of AGNs
by several gamma-ray space telescopes:
INTEGRAL~\footnote{http://sci.esa.int/integral/},
AGILE~\footnote{http://agile.rm.iasf.cnr.it/},
FERMI~\footnote{http://fermi.gsfc.nasa.gov/}, etc.
The gamma-ray flares were detected from more than 100 AGNs
including 3C~273, 4C~+21.35, 4C~454.3 and 1510-089,
by the Fermi Gamma-ray Space Telescope (GST)~\citep[see][for references]{kan+15}.
To understand the gamma-ray flares in the AGNs,
multiwavelength monitoring studies have been conducted
to investigate the correlation between emission at gamma-ray and other
wavelengths~\citep[e.g.,][]{mar+08,abd+10}.
The gamma-ray flares are correlated with 
the variability
in brightness, polarization, and morphology of the jets
at low energy regimes (e.g., radio, optical, etc.).
The correlation of the gamma-ray flares with multiwavelength variability,
in principle, implies co-spatiality of the gamma-ray flaring regions 
and the emission regions at other wavelengths.

\cite{mar+08} suggested that the gamma-ray flares from BL Lac may be generated
through inverse Compton scattering 
in the innermost region of the jets threaded by a helical magnetic field
or in a bright very long baseline interferometry (VLBI) core region
corresponding to a standing shock wave,
through compression by the shock front.
\cite{abd+10} reported that the gamma-ray flares correlated with the change of
optical polarization angle of 3C 279 provide evidence for 
the co-spatiality of the gamma-ray and optical emission regions
with a highly ordered jet magnetic field in a curved trajectory within the jet.
The different understandings of the cause and location of the gamma-ray flares
imply that either the gamma-ray flares are generated under various situations
or our understanding of the gamma-ray flares may still be uncertain.

Gamma-ray flares leading a delayed outburst at radio frequencies~\citep[e.g.,][]{jor+01}
motivated the conducting of many single-dish
and VLBI observations (or monitoring programs)
at centimeter (cm) and millimeter (mm) wavelengths~\citep{sav+02,LV03,lis+09,ojh+09,mar+11,ang+12,nag+13}.
Those VLBI monitoring programs are conducted only at a single frequency of $\le$43~GHz,
except for the TANAMI~\citep{ojh+09} at 8.4 and 22~GHz,
implying that those programs are not able to determine frequency-dependent properties
of the gamma-ray flaring AGNs on mas-scales,
including such properties as spectral index variation, 
time-dependent multiwavelength light curves of mas-scale structures, etc.,
which may be very important to investigate the correlation of the gamma-ray flares
with both the optically thin and thick emissions of the jets. 

The Korean VLBI Network (KVN) has enabled us to regularly
obtain mas-scale images of gamma-ray bright
AGNs at cm-to-mm wavelengths, leading to
a key science program (KSP) of the KVN, 
the Interferometric Monitoring of Gamma-ray Bright AGNs (iMOGABA),
which is a VLBI monitoring program for about 30 gamma-ray bright AGNs
using the KVN at multifrequency bands (22-129~GHz).
The iMOGABA initially started in 2012 December
by conducting monthly one 24-hr VLBI observation in the 22, 43, 86, and 129~GHz bands.
The iMOGABA has been selected as a KVN KSP since 2015 March,
aiming at studying the origins of the gamma-ray flares of the AGNs. 
We selected a total of 34 radio-loud AGNs,
as summarized in Table~\ref{t1},
of which 30 sources are gamma-ray bright AGNs with their flux densities
of $>6\times10^{-10}$~ph~cm$^{-2}$~s$^{-1}$,
including 24 sources monitored by the Fermi Gamma-ray Space Telescope
using the Large Area Telescope on board.
The selected sources consist of 24 quasars, 7 BL Lacs, and 3 radio galaxies.

In this paper,
the first representative paper from the iMOGABA program,
we report the results of a single-epoch VLBI observations
of the iMOGABA.
In Section~\ref{obs_analysis}, the observations and data analysis of the single-epoch VLBI session used in this paper is described.
Then, we present the results of the single-epoch observations in Section~\ref{results};
in Section~\ref{dis}, we present our discussion.
Finally, in Section~\ref{sum}, we summarize what we have found.


\section{Observations and data analysis\label{obs_analysis}}

\subsection{Observations\label{obs}}

We observed a total of 34 radio-loud AGNs
in the 22~GHz, 43~GHz, 86~GHz, and 129~GHz bands
on 2013 November 19 UT 02:00 - November 20 UT 01:53,
using the KVN which consists of three identical 21-m radio telescopes~\citep{lee+11}:
KVN Yonsei (KY), KVN Ulsan (KU), and KVN Tamna (KT),
with baseline lengths of 305--476~km~\citep{lee+14},
and a simultaneous multifrequency observing system~\citep{han+08,han+13}.
All sources were observed with 1-8 scans of 5~min in length.
The observing frequencies were 21.700--21.764~GHz, 43.400--43.464~GHz,
86.800--86.864~GHz, and 129.300--129.364~GHz, and in left circular polarization (LCP).
The ratio of the starting frequencies is an integer so that
the frequency phase technique can be applied to faint sources~\citep{alg+15}.
The total bandwidth is 256~MHz and is divided
evenly for each frequency band.
In order to measure changes in the opacity of the atmosphere
during the observations,
we conducted sky tipping curve measurements every hour, i.e.,
we measured the system temperatures at 8 elevations:
18.21$^\circ$, 20.17$^\circ$, 22.62$^\circ$, 25.77$^\circ$, 30.00$^\circ$,
36.03$^\circ$, 45.58$^\circ$, and 65.38$^\circ$,
during the observations.
Cross-scan observations for target sources were conducted every scan
to correct antenna pointing offsets,
which are the main source of amplitude uncertainty in KVN observations.
The received signals
were digitized (i.e. 2-bit quantized) by digital samplers
in the antenna cabin
and transmitted through optical fibers to the observing building.
The transferred digital signals were requantized and
divided into 16 sub-bands (IFs) of 16~MHz wide by a digital filter bank,
and recorded by the Mark 5B system at a recording rate of 1024~Mbps.
Correlation of the data was performed with the DiFX software correlator
in the Korea-Japan Correlator Center~\citep{lee+15a}.

\subsection{Data analysis\label{da}}

The DiFX correlator produced the spectrum
of the cross-correlation function with resolution of 0.125~MHz
and accumulation period of 1~s.
Post-correlation processing was done using
the NRAO Astronomical Image Processing System (AIPS).
A standard AIPS procedure was applied.
An antenna-based fringe fitting was conducted using AIPS task FRING.
KVN Ulsan (KU) was selected as the reference antenna.
The 5~min long scans were split into solution intervals that were 30~s long.
Baseline-based fringe solutions (phase delays and delay rates)
were searched for individual IF at all frequency bands.
The baseline-based fringe solutions were used to determine
the antenna-based fringe solutions.

All sources were detected at lower frequencies (e.g., 22-86~GHz) with this
standard post-correlation processing. However, about a dozen of the observed sources 
were not detected in the 129~GHz band.
In order to detect the non-detected sources,
the FPT technique was applied, using
an automatic VLBI data reduction pipeline developed for the KVN observations.
The operation of the KVN Pipeline was described in detail
by \cite{hod+16}.
Thanks to the FPT technique, i.e., by transferring phase solutions
from lower frequencies to higher 
as described by \cite{rio+11} and \cite{rio+14},
sensitivity at high frequency was improved,
and all of the non-detected sources were detected in the 129~GHz band.

Amplitude calibrations were performed based
on system temperatures measured during observations at each station
and on elevation dependence of the antenna gains provided.
The system temperatures were corrected for the atmospheric opacity
estimated from the sky tipping curve measurements conducted every hour.
We expect that thanks to the good quality of
the system temperatures, antenna gains, and atmospheric opacity for KVN stations,
the uncertainty of the amplitude calibration is witin 10-30\% at 22-129~GHz bands.
Re-normalization of the fringe amplitudes was made 
to correct for the amplitude distortion due to quantization,
and the quantization and re-quantization losses were
corrected~\citep{lee+15b}.

\subsection{Imaging with calibrated data\label{imaging}}

The phase- and amplitude-calibrated data were used
to produce CLEAN contour maps of the sources
via the Caltech DIFMAP software~\citep{she+94}.
The calibrated data were averaged in frequency,
and the data at band edges were chopped to avoid flux
loss due to bandpass shape.
The {\it uv}-data were also averaged in time over 30 seconds in the 22, 43, and 86~GHz bands,
and over 10 seconds in the 129~GHz band,
taking into account coherence time at each frequency bands,
and expecting a decoherence loss of upto 30\% at 86-129GHz bands.
Some outliers in the $uv$-data were flagged
by investigating the visibility amplitude. 

The averaged and edited {\it uv}-data were then
fitted using a single circular two-dimensional Gaussian model, and
the visibility phase was self-calibrated with the model.
After removing the single Gaussian model, a combination of CLEAN point source
models was applied.
The visibility phase was again self-calibrated with the CLEAN model
to produce the final CLEAN contour maps of the detected sources, when available.
The visibility amplitude was not modified.

\begin{figure*}[!t]
\epsscale{0.9}
\plotone{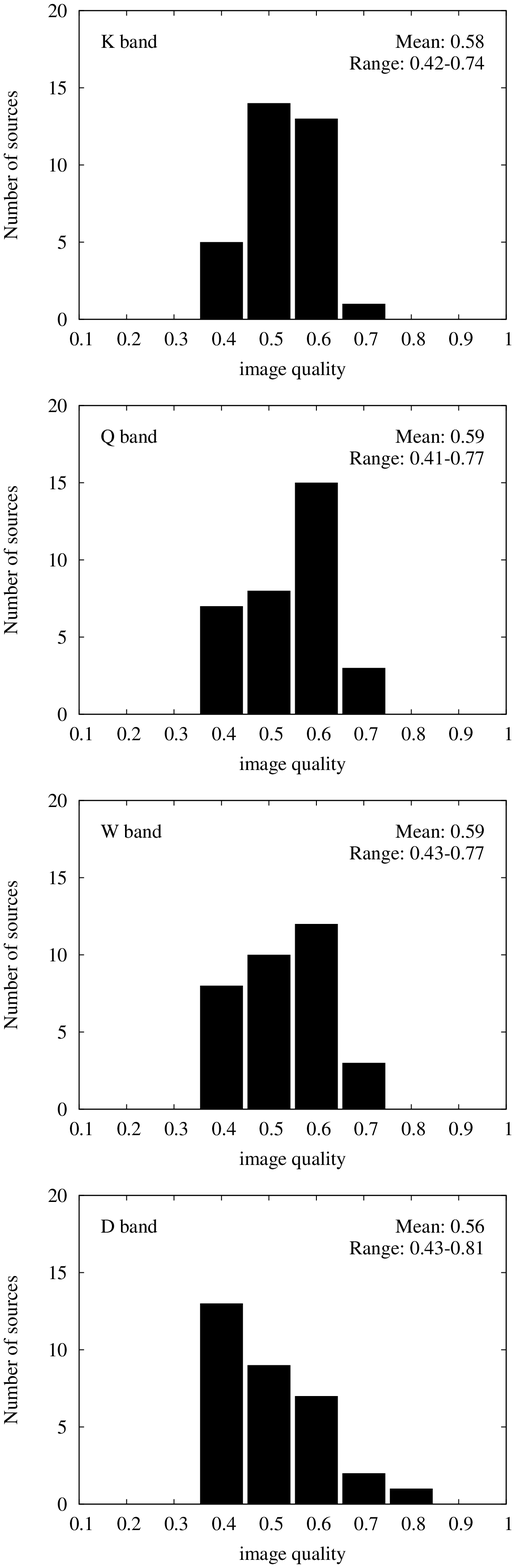}
\caption{Distributions of the image quality factor $\xi_{\rm r}$.
The mean and range of the distributions are presented.
\label{fig-hist-q}}
\end{figure*}

The residual noise in the final CLEAN map was investigated
to evaluate the quality of the image,
taking into account the assumption
that the residual noise of the properly CLEAN image is random noise,
and hence the amplitude of each pixel in the residual map should have
a Gaussian distribution.
So, the noise in the final image can be expressed quantitatively 
by the ratio of image noise rms to its mathematical expectation,
$\xi_{\rm r} = s_{\rm r} / s_{\rm {r,exp}}$,
where $|s_{\rm r}|$ is the maximum absolute flux density in the residual image
and $|s_{\rm r,exp}|$ is the expectation of $s_{\rm r}$. 
For Gaussian noise with a zero mean,
the expectation of $s_{\rm r}$ is
$|s_{\rm {r,exp}}| = \sigma_{\rm r} { \left[ \sqrt{2} \ln{\left( \frac{N_{\rm pix}} {\sqrt{2\pi}\sigma_{\rm r}} \right) }\right]}^{1/2}$,
where $N_{\rm pix}$ is the total number of pixels in the image,
and $\sigma_{\rm r}$ is the image noise rms.
If $ \xi_{\rm r} \rightarrow 1 $, the residual noise approaches Gaussian noise; 
if $ \xi_{\rm r} > 1 $, not all the structure has been adequately recovered; 
if $ \xi_{\rm r} < 1 $, the image model has
an excessively large number of degrees of freedom~\citep{lob+06}.
The values of $\xi_{\rm r} $ of the images obtained in this paper
are 0.42-0.74 at 22~GHz,
0.41-0.77 at 43~GHz,
0.43-0.75 at 86~GHz,
and 0.43-0.81 at 129~GHz,
as shown in Figure~\ref{fig-hist-q}
and summarized in Table~\ref{t2},
implying that the images adequately represent the structure 
detected in the visibility data.

\subsection{Model fitting and estimating parameters\label{modelfit}}

Circular two-dimensional Gaussian models were used to fit
the final CLEAN images, yielding the general fit parameters:
{\it$S_{\rm tot}$} (total flux density), {\it $S_{\rm peak}$} (peak flux density), 
{\it $\sigma_{\rm rms}$} (post-fit rms), $d$ (size), 
$r$ (radial distance, for jet components), 
$\theta$ (position angle, measured for jet components, with respect to the location of 
the core component), as summarized in Table~\ref{t3}.
The uncertainties of the fit parameters 
were estimated, based on the signal-to-noise ratio
of each component detection, by adopting an analytical first order approximation
given by~\cite{fom99}:
$\sigma_{\rm peak} = \sigma_{\rm rms}{\left({1+\frac{S_{\rm peak}}{\sigma_{\rm rms}}}\right)}^{1/2}$,
$\sigma_{\rm tot} = \sigma_{\rm peak}{\left({1+\frac{S_{\rm tot}^2}{S_{\rm peak}^2}}\right)}^{1/2}$,
$\sigma_{d} = d\frac{\sigma_{\rm peak}}{S_{\rm peak}}$,
$\sigma_{r} = \frac{1}{2}\sigma_{d}$,
and
$\sigma_{\theta} ={\rm atan}\left(\frac{\sigma_{r}}{r}\right)$,
where
$\sigma_{\rm peak}$,
$\sigma_{\rm tot}$,
$\sigma_{\rm rms}$,
$\sigma_{d}$, 
$\sigma_{r}$,
and
$\sigma_{\theta}$
are the uncertainties of
peak flux density,
total flux density,
post-fit rms,
size, 
radial distance,
and
position angle of a component, respectively. 
A component was considered to be unresolved when the fitted component size $d$
is smaller than its minimum resolvable size~\citep{lob05} as given by
$d_{\rm min}=\frac{2^{1+\beta/2}}{\pi}{\left[{\pi ab\ln2\ln{\frac{\rm S/N}{{\rm S/N}-1}}}\right]}^{1/2}$,
where {\it a} and {\it b} are the axes of the restoring beam, 
S/N is the signal-to-noise ratio, and $\beta$ is the 
weighting function, which is 0 for natural weighting or 2 
for uniform weighting.
If $d<d_{\rm min}$, the uncertainties were estimated according to $d=d_{\rm min}$.

The total flux density and measured size of a Gaussian component leads to 
a simple estimation of the rest frame brightness temperature $T_{\rm b}$
of the given emission region represented by the Gaussian model as
$T_{\rm b} = \frac{2\ln{2}}{{\pi}{k}}\frac{S_{\rm tot}\lambda^2}{d^2}(1+z)$,
where $\lambda$ is the wavelength of observation, $z$ is the redshift, 
and $k$ is the Boltzmann constant. If $d < d_{\rm min}$,
then the lower limit of $T_{\rm b}$ is obtained using $d=d_{\rm min}$.

\section{Results\label{results}}

Out of 34 sources, 32 sources were detected and imaged at all frequency bands
with or without a frequency phase transfer technique
that enabled us to detect and image 13 faint sources at 129~GHz,
except for 0218+357
which was detected for only one baseline at all frequency bands.
The first 129~GHz VLBI images were made for most of the sources in this project.

\begin{figure*}[!t]
\epsscale{2}
\plotone{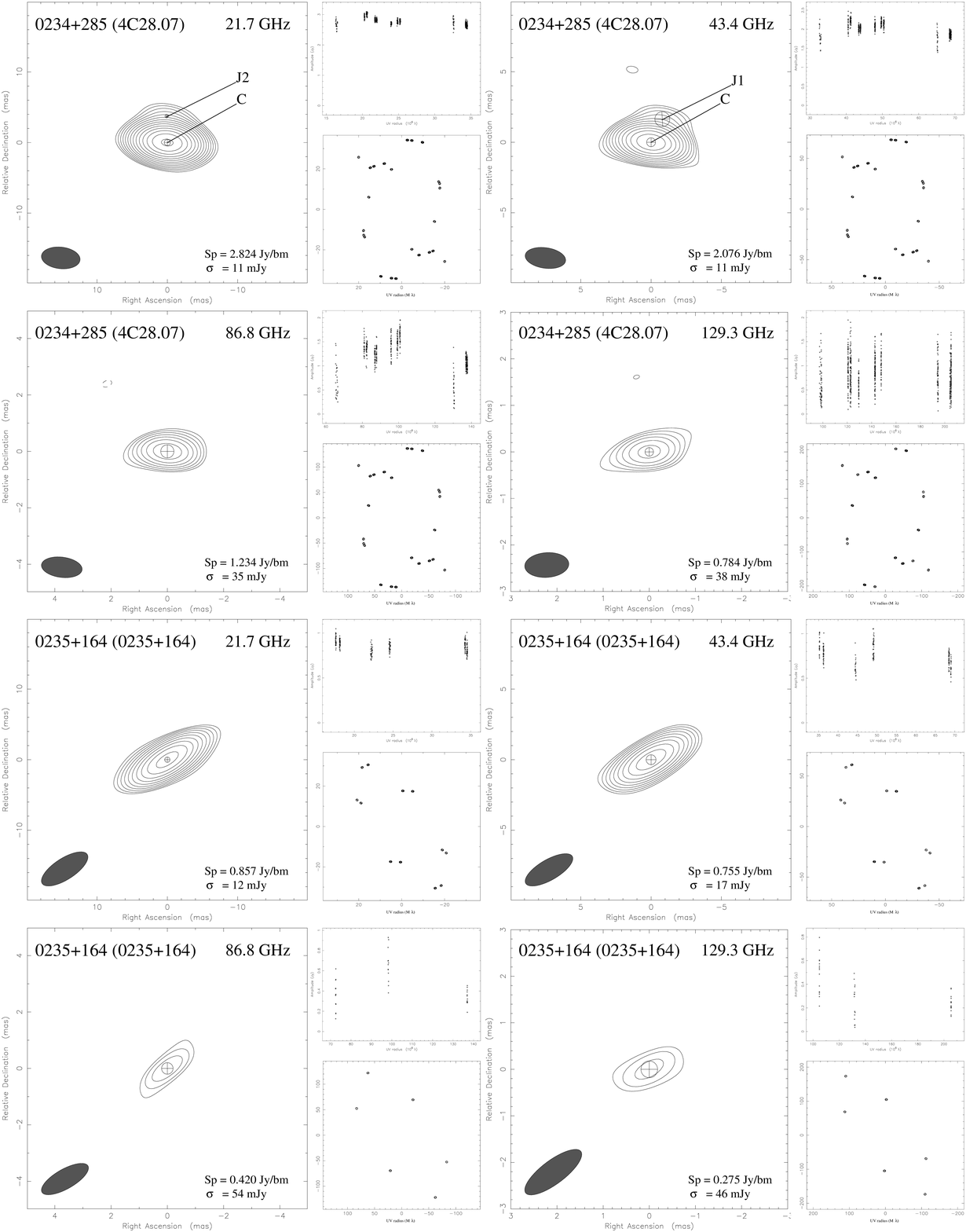}
\caption{
Contour maps of all sources with the distributions of
the {\it uv}-sampling and of the visibility amplitude against {\it uv}-radius. 
In the left panel,
a contour map of the CLEANed image is shown. 
The axes of the maps show the relative offset from the center of image in mas.
Minimum contour level is shown in the lower-right corner of each map.
The contours have a logarithmic spacing and are drawn at 
-1, 1, 1.4,...,$1.4^n$ of the minimum contour level. 
In the upper right panel,
the X-axis represents the visibility amplitude (correlated flux density)
in Jy, and the Y-axis shows the {\it uv}-distance in $10^6 \lambda$. 
The corresponding {\it uv}-sampling distribution is given in the lower right palen.
Image parameters of each image are summarized in Table~\ref{t2}.
\label{fig-image}}
\end{figure*}

In Figure~\ref{fig-image}, CLEAN contour maps for each source are
presented for the 22, 43, 86, and 129~GHz bands, when available.
A set of plots for each source at each frequency consists of  
a contour map in the left panel,
a plot of the corresponding visibility amplitude
against {\it uv}-radius in the top right panel,
and a plot of the corresponding {\it uv}-sampling distribution
in the bottom right panel.
The contour map is shown with the X- and Y-axes in units of mas.
The source name and the observing frequency are given at
the top of the map.
The FWHM of the restoring beam is drawn as a shaded ellipse on the map.
The peak flux density and rms noise level are identified
in the lower right corner of the map.
The contours are drawn with a logarithmic spacing at -1, 1, 1.4,...,$1.4^n$
of the lowest flux density level which corresponds to three times 
the rms noise level. 
Most contour maps are centered on their brightest emission region
(which is considered to be a VLBI core with fainter emission regions being
the corresponding jet component, except for 3C~84 and 4C+39.25),
but for some maps with extended source structure,
the map center is shifted to fit the map in the box.
The plot of the visibility amplitude, at the top right of each set,
is drawn as a function of the {\it uv}-radius,
i.e., the length of the baseline used to obtain the corresponding visibility
point in units of $10^6\lambda$,
where $\lambda$ is the observing wavelength. 
The visibility amplitude (i.e., the correlated flux density)
is shown in units of Jy.
The {\it uv}-sampling distribution, at the bottom right of each set,
shows the distribution of the visibility in the {\it uv}-plane,
whose range corresponds to that of the {\it uv}-radius.
For the 12 sources
(0235+164,
0528+134,
0735+178,
0827+243,
0836+710,
1127-145,
1156+295,
1222+216,
1343+451,
1611+343,
and
3C~286)
the FPTed images, i.e., the images obtained by the FPT technique, are presented
for the 129~GHz band.
In particular, the 43 and 86~GHz images for 3C~286 are also FPTed.

The KVN observations for this paper were done without the phase-referencing technique,
i.e., without relative (or absolute) astrometric information of core or jet components
with respect to calibrators. So, we considered a brightest emission region to be a VLBI core
in each image and the others the jet, following a general strategy for identifying VLBI cores.
However, we found that for some cases, maybe rare but well-known, the brightest component may
not be the core. For example, 3C~84 (0316+413),
the J3 component is brighter than the core (C) at 43 and 86~GHz.
At 43~GHz, in particular, the J3 component was centered and the core was located in the northern
area in the beginning of the imaging process due to its brightness.
Considering the spectral index of the core (C) to be flat, we identified
the northern component as a core, and shifted the map to south.
Another exceptional source is 4C~$+$39.25 (0923+392).
The eastern component (J1) is brighter than the western component (C) and was located 
in the map center in the imaging process.
Since, however, the fainter component is usually identified as the core~\citep[e.g.,][]{nii+14},
we considered the fainter component as the core and shifted the map to have the core in the center.

Parameters of the contour maps presented in Figure~\ref{fig-image} are summarized
in Table~\ref{t2}.
For each image, 
Table~\ref{t2} lists the source name, the observing frequency band, 
the parameters of the restoring beam (the size of the major axis $B_{\rm maj}$,
the minor axis $B_{\rm min}$, and the position angle of the beam $B_{\rm PA}$),
the total flux $S_{\rm KVN}$,
the peak flux density $S_{\rm p}$,
the off-source RMS $\sigma$,
the dynamic range of the image $D$,
and the quality $\xi_{\rm r}$ of the residual 
noise in the image.

The parameters of each model-fit component are listed in Table~\ref{t3}:
the total flux $S_{\rm tot}$,
peak flux density $S_{\rm peak}$,
size $d$,
radius, $r$ (only for jet components),
position angle $\theta$ (only for jet components),
and measured brightness temperature $T_{\rm b}$. 
In cases of multiple components (i.e., VLBI core and jet components) fitted,
the parameters of the core component are followed by those of the jet components.
The estimated uncertainties are given as in Section~\ref{modelfit}.
The upper limits of size $d$, 
and the lower limits of brightness temperature $T_{\rm b}$
are in italics with flag symbols.

\section{Discussion\label{dis}}

\subsection{Source morphology on mas scales}

Many of the target sources are resolved on mas scales,
yielding a core-jet structure.
Out of 33 sources imaged, 22 sources show a core-jet structure
at at least one of the observing frequency bands.
Most of the sources with core-jet structure
are imaged to show a single jet component, except for 
0316+413 (3C~84),	 
1226+023 (3C~273B),	 
1611+343,   		 
1641+399 (3C~345),       
2230+114 (CTA102),       
and 2251+158 (3C~454.3), 
which show more than one jet component at at least one frequency band
on the scales of 1-11~mas,
contributing to the CLEAN flux density of the sources by $<$50\%,
as summarized in Table~\ref{t3}.
Although many of the target sources are resolved,
they are highly core-dominated in flux, i.e.,
the VLBI core flux density ($S_{\rm c}$) dominates the synchrotron emission
on the mas-scales ($S_{\rm KVN}$),
as shown in Figure~\ref{fig-hist-ratio}.
Some of the very compact sources show a ratio of $S_{\rm c}/S_{\rm KVN}$
larger than unity due to uncertainty of the CLEAN flux and of the model fitting.

The distributions of angular size for the core components of the imaged sources 
were investigated.
As introduced in Section~\ref{modelfit},
a core component was considered to be unresolved when the fitted core size
was smaller than its minimum resolvable size,
yielding the upper limit of the angular size.
The numbers of unresolved cores are 14 (22~GHz), 3 (43~GHz), 4 (86~GHz),
and 5 (129~GHz) out of 33 imaged sources.
Excluding the upper limits, the mean (and median) values of the measured core angular sizes are 
1.135~mas (0.912~mas), 0.586~mas (0.591~mas), 0.411~mas (0.438~mas), and 0.232~mas (0.179~mas)
in the 22, 43, 86, and 129~GHz bands, respectively,
implying that the resolved core sizes become smaller at higher frequencies.

\subsection{Brightness temperature}\label{tb}
Figure~\ref{fig-hist-tb} shows the brightness temperatures of the VLBI cores,
including the lower limits (in grey),
which were obtained for the unresolved cores (see Section~\ref{modelfit}).
Excluding the limits, the measured brightness temperatures are in the ranges of
$2.1\times10^{7}$~K - $7.5\times10^{8}$~K,
$2.1\times10^{7}$~K - $3.8\times10^{10}$~K,
$1.4\times10^{8}$~K - $3.2\times10^{10}$~K,
and
$7.7\times10^{8}$~K - $1.2\times10^{11}$~K,
at 22, 43, 86, and 129 GHz, respectively.
The mean values of the measured brightness temperature are
$1.7\times10^{9}$~K,
$4.2\times10^{9}$~K,
$6.0\times10^{9}$~K,
and $1.7\times10^{10}$~K
in the 22, 43, 86, and 129~GHz bands, respectively,
implying that the brightness temperatures of the sources become higher at higher frequency.

These results are different from those reported
by \cite{lee13}, \cite{lee14}, and \cite{lee+16}:
the brightness temperatures of the VLBI cores imaged at 86~GHz
with the Global Millimeter VLBI Array (GMVA) are lower than 
those obtained at 2-15~GHz with the global VLBI array.
The mean values of the brightness temperatures for the VLBI cores
in \cite{lee+16} are
$8.9\times10^{11}$~K
and $1.3\times10^{11}$~K
at 15 and 86~GHz, respectively;
these values are higher than those we obtained in this work,
by factors of about 50 and 20, respectively, although the source samples and their sizes
are different from each other.
This difference may be due to the spatial resolution difference
between the global VLBI arrays, which have a maximum baseline of about 10000~km,
and the KVN, with
its maximum baseline of about 500~km.
However, the lower brightness temperatures at lower frequencies (e.g., 22~GHz) 
obtained with the KVN cannot be fully explained by the resolution difference.
One explanation may be that the sources resolved at 22~GHz with the KVN
are largely different from those resolved with the global array, since half of the cores
in our sample were unresolved.
Another explanation may be 
that the brightness temperature measurements at lower frequencies are affected
partly by an instrumental beam blending effect:
the core region may be diluted by jet emission~\citep[see e.g.,][and references therein]{lee+16}.
The core blending effect would make larger
both the observed core flux density and the size. 
In this case of the 22~GHz observations with the KVN, 
the increasing factor of the observed core size $d$
is much larger than that of the observed core flux density $S_{\rm tot}$,
and hence the brightness temperature ($\propto S_{\rm tot}/d^2$) becomes lower than its true value.


\subsection{Source spectra}

The 129~GHz images presented in this paper are the first results for most of the radio
loud AGNs in our sample. Moreover, the simultaneous multifrequency images obtained
at four frequency bands (22-129~GHz) are also the first results for the radio-loud AGNs,
enabling us to investigate mas-scale spectral information for the radio loud AGNs
free from any variability of source. Here we discuss the spectral
properties of the target sources, taking into account the CLEAN flux densities
obtained on mas-scales.

The CLEAN flux densities $S_{\rm KVN}$ of the target sources obtained with the KVN are
in the ranges of 
0.43-28~Jy,
0.32-21~Jy,
0.18-11~Jy,
and 0.35-8.0~Jy
in the 22, 43, 86, and 129~GHz bands, respectively,
as shown in Figure~\ref{fig-hist-CLEAN}.
The distributions of the CLEAN flux density
peak at around 2~Jy (22 and 43~GHz bands) and
1~Jy (86 and 129~GHz bands);
these values show that most of the sources are brighter
than 0.1~Jy at all frequencies.
The mean values of the CLEAN flux density are
4.7~Jy (21.7~GHz), 3.8~Jy (43.4~GHz), 2.5~Jy (86.8~GHz), and 1.6~Jy (129.3~GHz),
implying that the sources become fainter at higher frequency on mas-scales.


To investigate the spectral properties of the target sources,
three models were used to fit it to the CLEAN flux density data:
the power law,
the broken power law,
and the curved power law,
as given by
\begin{equation}
\label{eqn:plaw}
S=a \nu^{\alpha},
\end{equation}
where $S$ is the CLEAN flux density in Jansky,
$\nu$ is the observing frequency in GHz,
$a$ is constant in Jansky,
and $\alpha$ is the spectral index for the power law,
\begin{equation}
\label{eqn:blaw}
S=b \left\{\left(\frac{\nu}{\nu_{\rm c}}\right)^{-\alpha_{\rm L}n}+\left(\frac{\nu}{\nu_{\rm c}}\right)^{-\alpha_{\rm H}n}\right\}^{-1/n},
\end{equation}
where
$b$ is constant in Jansky,
$\nu_{\rm c}$ is a break frequency of the broken power law,
$\alpha_{\rm L}$ and $\alpha_{\rm H}$ are the spectral indices at lower and higher
frequencies than the break frequency,
and $n$ is a numerical factor governing the sharpness of the break,
here fixed to be 100 in order to make the break very sharp;
and,
\begin{equation}
\label{eqn:claw}
S=c_1 \left(\frac{\nu}{\nu_{\rm r}}\right)^{\alpha +c_2 {\rm ln}(\nu/\nu_{\rm r})},
\end{equation}
where
$c_1$ (in Jansky) and $c_2$ are constant,
$\nu_{\rm r}$ is a reference frequency,
and $\alpha$ is the spectral index at $\nu_{\rm r}$.
As a first step, we attempt to fit the power law to the data.
When the power law does not fit to the data, we use the curved power law.  
For 7 sources, the power law and curved power law do not fit well to the data
and, hence, a broken power law is used to fit to the data for those sources.
The CLEAN flux spectra of the 32 sources (except for 3C~286) are shown
in Figure~\ref{fig-spind} with their best fitting model and
the best fitting parameters:
the spectral index $\alpha$ for the power law,
the cut-off (turnover) frequency $\nu_{\rm c}$,
the spectral indices at lower ($\alpha_{\rm L}$) and higher ($\alpha_{\rm H}$) frequencies
than the break frequency,
obtained from the broken power law,
the cut-off (turnover) frequency $\nu_{\rm c}$,
the peak flux density $S_{\rm m}$, and
the spectral indices between
22-43~GHz (KQ), 43-86~GHz (QW), and 86-129~GHz (WD),
obtained from the curved power law. 
Out of 32 sources, only 5 sources show the spectra well fitted with the power law;
27 sources have spectra with either a break or a curvature. For the 32 spectra shown
in Figure~\ref{fig-spind},
we used the best fitting model to obtain spectral indices in the KQ-band, QW-band, and WD-band.
Figure~\ref{fig-hist-spind} shows the distributions of the spectral indices,
with mean (median) values of -0.40 (-0.42), -0.62 (-0.66), and -1.00 (-0.99)
in the KQ, QW, and WD bands, respectively,
indicating that 
many sources become optically thin at higher frequency (e.g., in the 86 and 129 GHz bands).

%

\subsection{VLBI core size and jet geometry}

In the VLBI observations, the location of the $\tau=1$ surface of the AGN thick core depends
on the frequency according to the relation $r\propto\nu^{-1/k_r}$, leading to the well-known
`core-shift' effect~\citep[see, e.g.][]{lob98,hir05,OG09,alg+12}.
This implies that, at various observing frequencies, the VLBI core actually probes
different distances from the SMBH. At the same time, if we consider the VLBI core
as the innermost upstream regions of the unresolved jet, this suggests that measuring
the core size at different frequencies can help us to investigate the innermost jet size
at different locations, via the core-shift and, hence, the jet's geometry.
This approach has already been successfully applied in, for example, \cite{had+11} and \cite{alg+16}.
However, the former considered only one source,
and the latter considered observations spanning over several years.
The iMOGABA does not have any of these drawbacks and can consider a whole sample
of simultaneous multi-frequency observations.
 
In Figure~\ref{fig-geometry} we show a representative subset of sources
to which this method has been applied. Here we plot the core radius (1/2 of the size)
against the inverse of the frequency, which is proportional to the distance from the SMBH.
We performed an exponential fit following $(size)\propto(distance^{\epsilon})$
to find the inferred jet geometry
($\epsilon=1$ conical; $\epsilon=1/2$ parabolic; c.f. \cite{ghi+85}).
In general, we can distinguish three different trends:
i) all points seem to be reasonably well aligned and fit to a jet size increasing
with distance (left column), ii) a certain convex trend is found for the data (middle column)
or iii) other peculiar patterns are found (right column).
In the following we briefly discuss these cases.

There are several cases in which the data are well fitted by an exponential function,
in agreement with the discussion above. The largest group, however, shows an increase of
the core size, which does not follow the expected exponential trend,
but rather seems to show a size saturation threshold at lower frequencies.
One possibility is that, due to the large scatter and limited data in the {\it uv}-plane,
a small offset in the amplitude in one scan may cause the fitted Gaussian size to be slightly
different, especially in the case of compact sources, which are common in the iMOGABA sample.

In some cases, the fit indicates a jet with a size that roughly constant or decreasing with size;
the jet may even show a possibly oscillatory pattern.
Whilst this phenomenology could be true for a reduced number of sources
in very specific conditions for a given time,
we believe that this is not the case here.
Instead, we suggest that this behavior is rather due to other effects such as jet blending,
the existence of various unresolved components, or extremely large viewing angles.
As for the case mentioned above, the lack of a significant amount of data in the {\it uv}-plane
may also contribute to a certain extent
to the misidentification of some core sizes;
thus, care should be taken. Averaging of the data through various epochs may partially
solve this issue. This will be discussed in a forthcoming paper.
Further discussion can also be found in \cite{alg+16}.

\section{Summary\label{sum}}

An interferometric monitoring program for gamma-ray bright AGNs was launched
and, 
by conducting monthly multifrequency VLBI observations of a total of 34 radio loud AGNs,
continues to aim at revealing the origins of gamma-ray flares often detected
in AGNs.
One of the observing sessions in the program was conducted
in 2013 on November 19 and 20,
yielding 32 sources detected and imaged at all frequency bands
with or without use of frequency phase transfer technique.
The first 129~GHz VLBI images were made from most of the sources,
establishing a 129~GHz VLBI image database with a size of 32 sources.
In addition to the CLEAN images, we used two-dimensional circular Gaussian
models to fit the CLEAN images and obtained the VLBI core sizes and brightness temperatures.

We find that 
the 22 sources show the core-jet structure
at at least one of the observing frequency bands.
Most of the sources with the core-jet structure
are imaged to show a single jet component, except for 
six sources with multiple jet components
on the scales of 1-11~mas,
contributing to the CLEAN flux density of the sources by $<$50\%.
Although many of the target sources are resolved,
they are highly core-dominated in flux.
We also find that the resolved core sizes become smaller at higher frequencies,
with the mean values of the measured core angular sizes
of 1.135~mas, 0.586~mas, 0.411~mas, and 0.232~mas
in the 22, 43, 86, and 129~GHz bands, respectively.

The mean values of the measured brightness temperature are
$1.7\times10^{9}$~K,
$4.2\times10^{9}$~K,
$6.0\times10^{9}$~K,
and $1.7\times10^{10}$~K
in the 22, 43, 86, and 129~GHz bands, respectively,
implying that the brightness temperatures of the sources become higher at higher frequency.
We find that these brightness temperatures are lower than those obtained with global
VLBI arrays, indicating that the brightness temperature measurements
are affected by the low spatial resolution of the KVN observations
and by the instrumental beam blending effect, especially in the 22~GHz band.

Thanks to the simultaneous multifrequency VLBI observations, 
we are able to investigate the spectral properties free from any source variability.
We find that the sources become fainter at higher frequency,
yielding optically thin spectra at mm wavelengths.
More sophisticated modelling studies for spectral energy distribution with data in broader wavelength bands will be presented in forthcoming papers on individual sources in the iMOGABA program.

\acknowledgments
We would like to thank the anonymous referee for important comments
and suggestions which have enormously improved the manuscript.
We are grateful to all staff members at KVN
who helped to operate the array and to correlate the data.
The KVN is a facility operated by
the Korea Astronomy and Space Science Institute.
The KVN operations are supported
by KREONET (Korea Research Environment Open NETwork)
which is managed and operated
by KISTI (Korea Institute of Science and Technology Information).
This work was supported by the National Research Foundation of Korea(NRF) grant funded by the Korea government(MSIP) (No. NRF-2016R1C1B2006697).
Dae-Won Kim and Sascha Trippe acknowledge support from the National Research Foundation of Korea (NRF) via grant NRF-2015-R1D1A1A-01056807.
Jongho Park acknowledges support via NRF PhD fellowship grant 2014H1A2A1018695.
Guang-Yao Zhao acknowledges the support from Korea Research Fellowship Program through the National Research Foundation of Korea(NRF) funded by the Ministry of Science, ICT and Future Planning (NRF-2015H1D3A1066561).

{\it Facilities:} \facility{KVN}



\clearpage

\clearpage

\setcounter{figure}{1}
\begin{figure*}[!t]
\epsscale{2}
\plotone{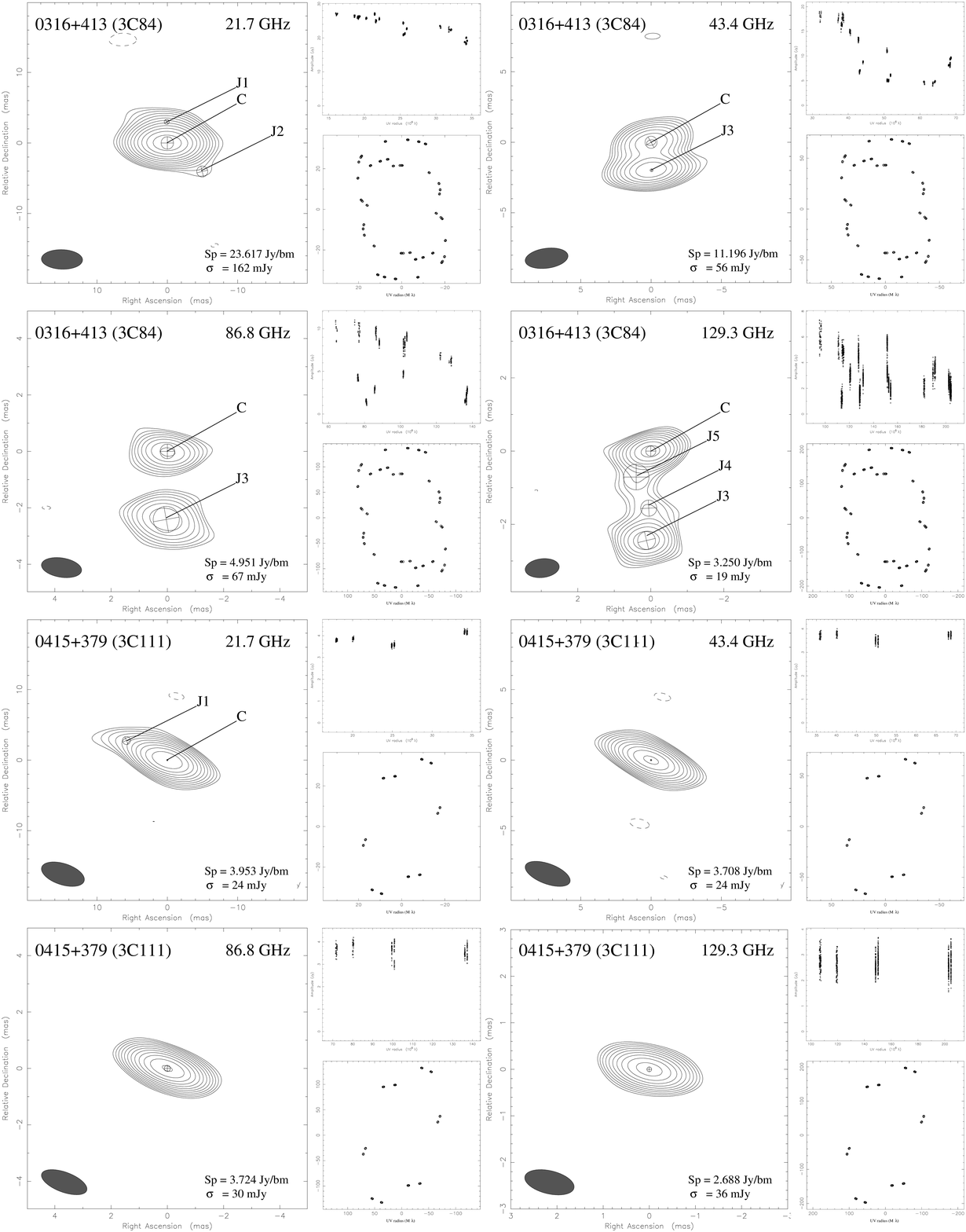}
\caption{{\it CLEAN images (continued)}}
\end{figure*}
\setcounter{figure}{1}
\begin{figure*}[!t]
\epsscale{2}
\plotone{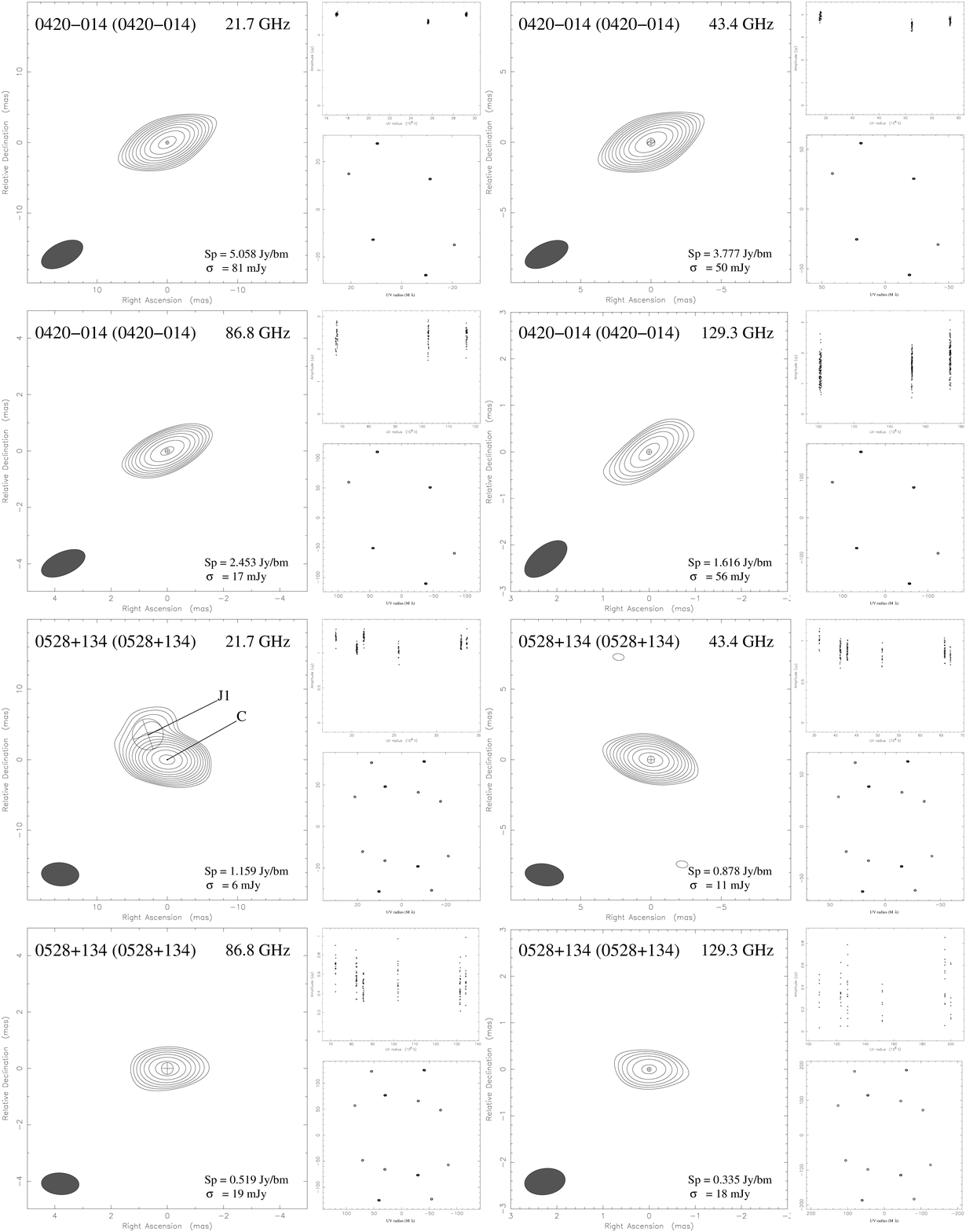}
\caption{{\it CLEAN images (continued)}}
\end{figure*}
\setcounter{figure}{1}
\begin{figure*}[!t]
\epsscale{2}
\plotone{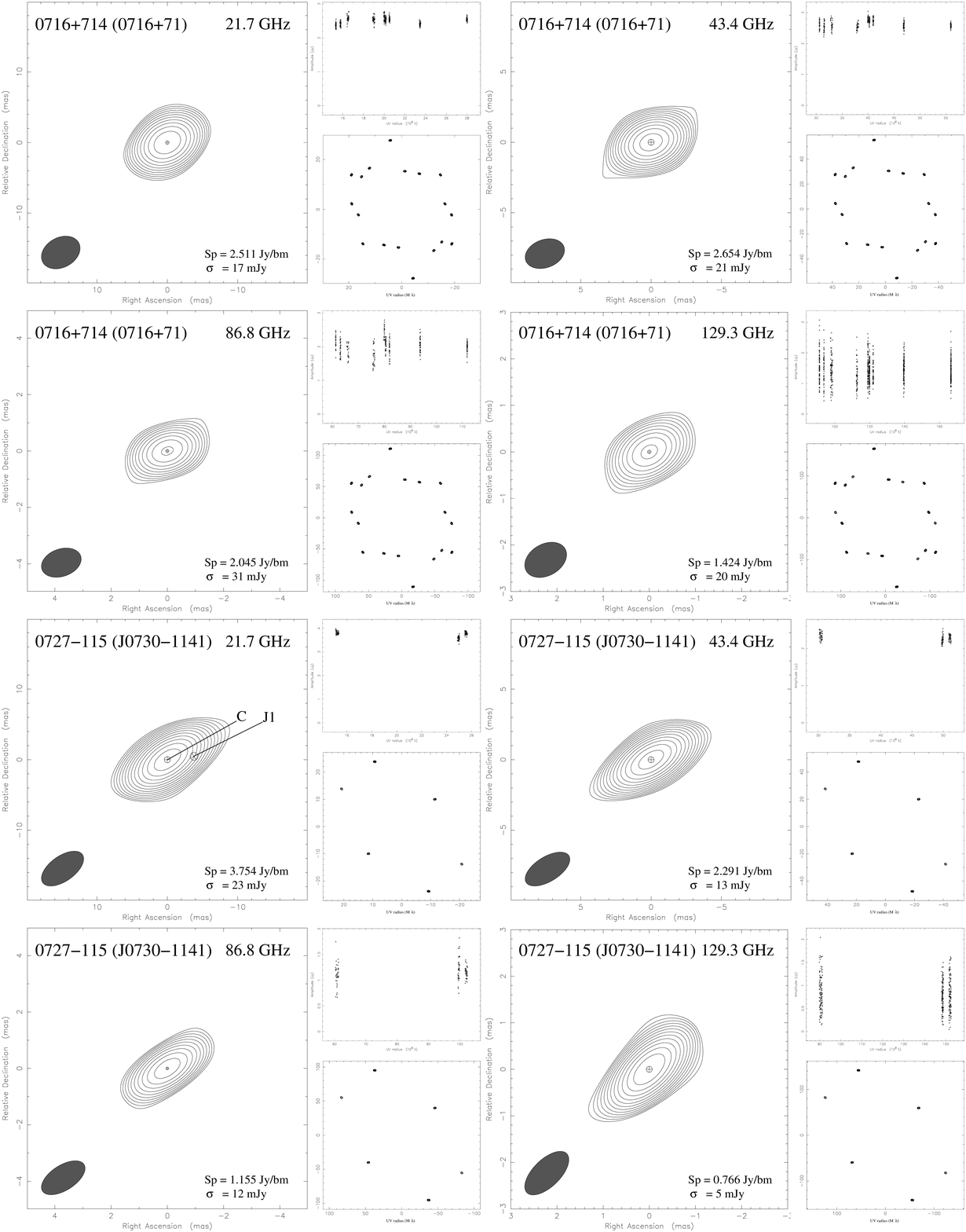}
\caption{{\it CLEAN images (continued)}}
\end{figure*}
\setcounter{figure}{1}
\begin{figure*}[!t]
\epsscale{2}
\plotone{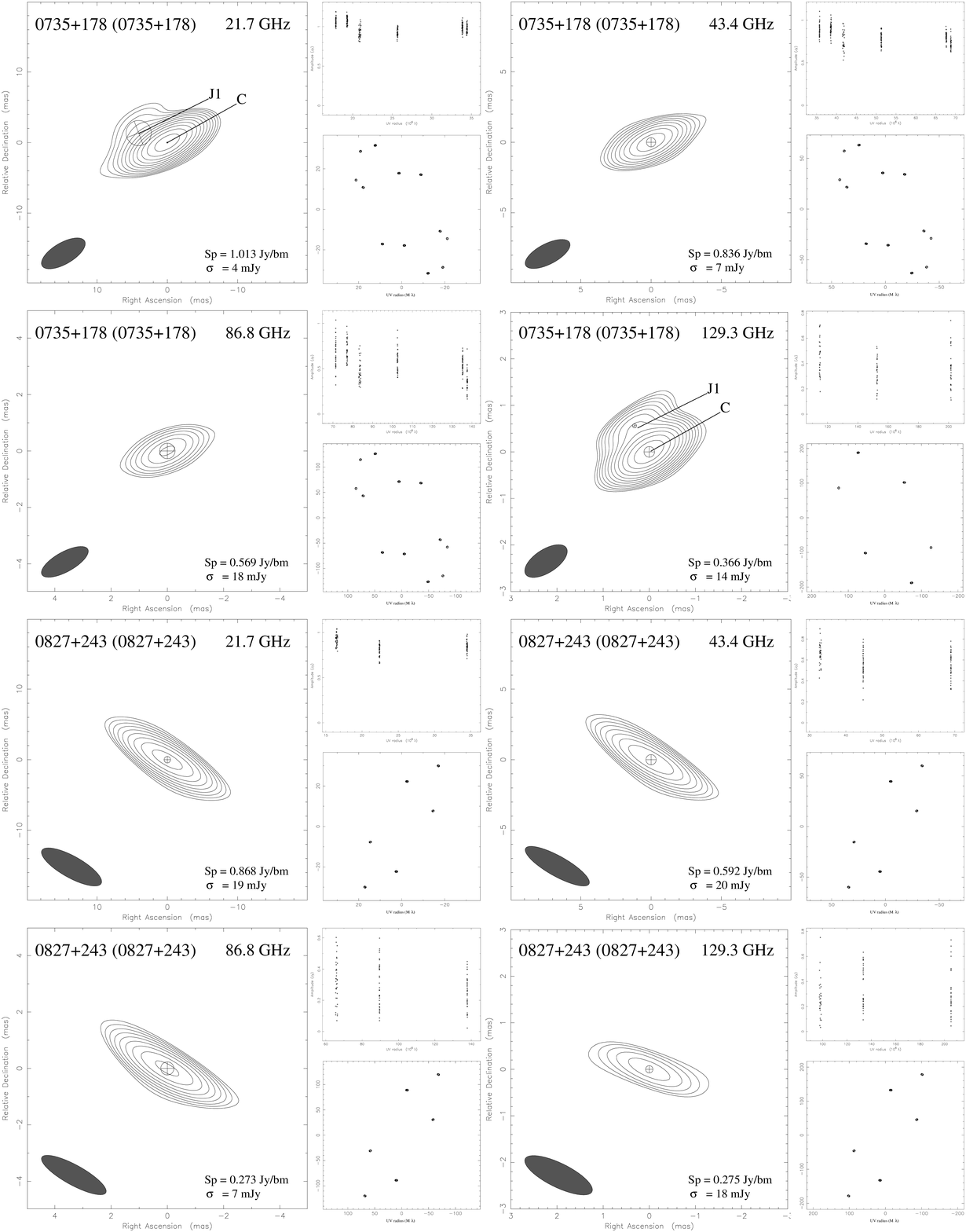}
\caption{{\it CLEAN images (continued)}}
\end{figure*}
\setcounter{figure}{1}
\begin{figure*}[!t]
\epsscale{2}
\plotone{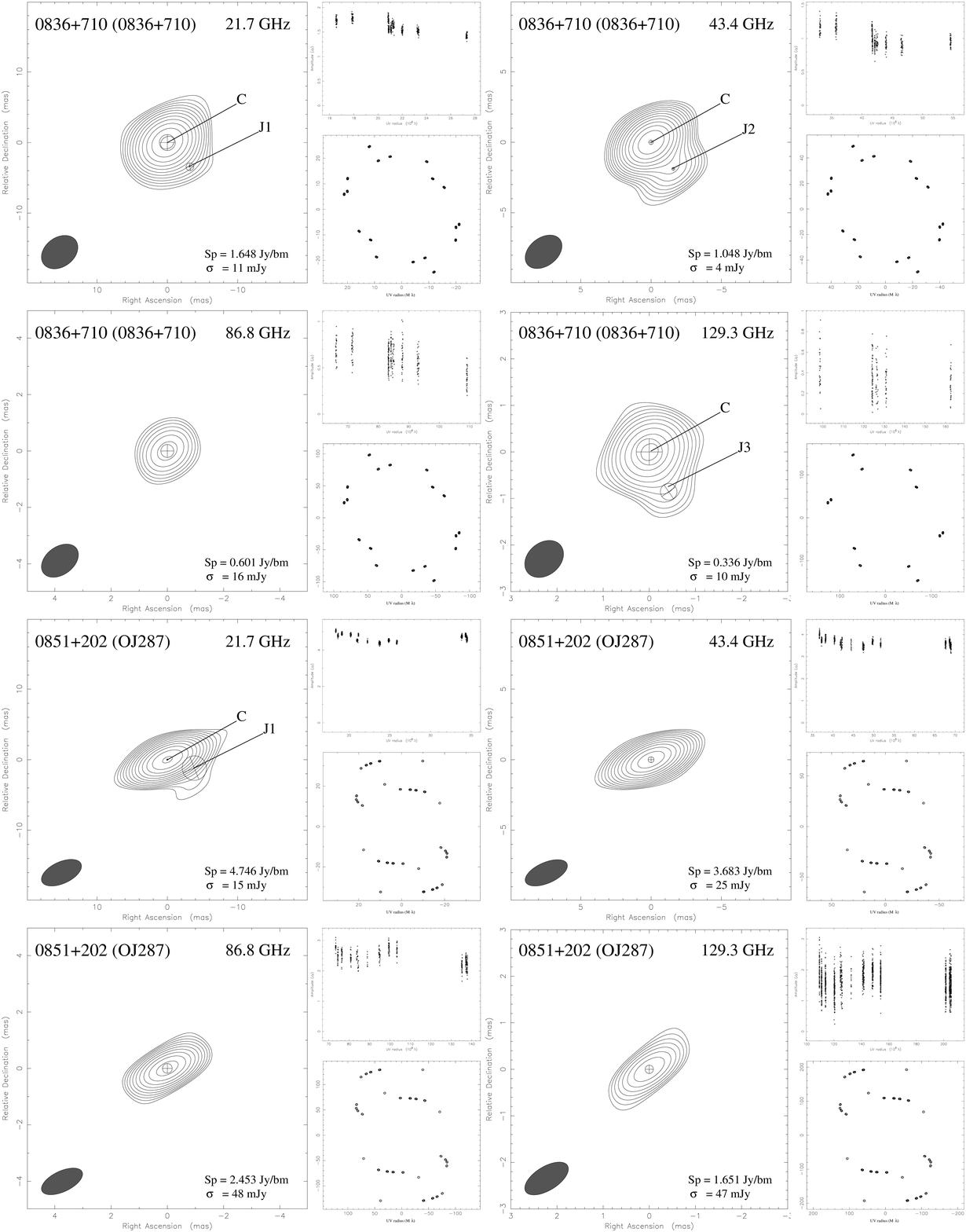}
\caption{{\it CLEAN images (continued)}}
\end{figure*}
\setcounter{figure}{1}
\begin{figure*}[!t]
\epsscale{2}
\plotone{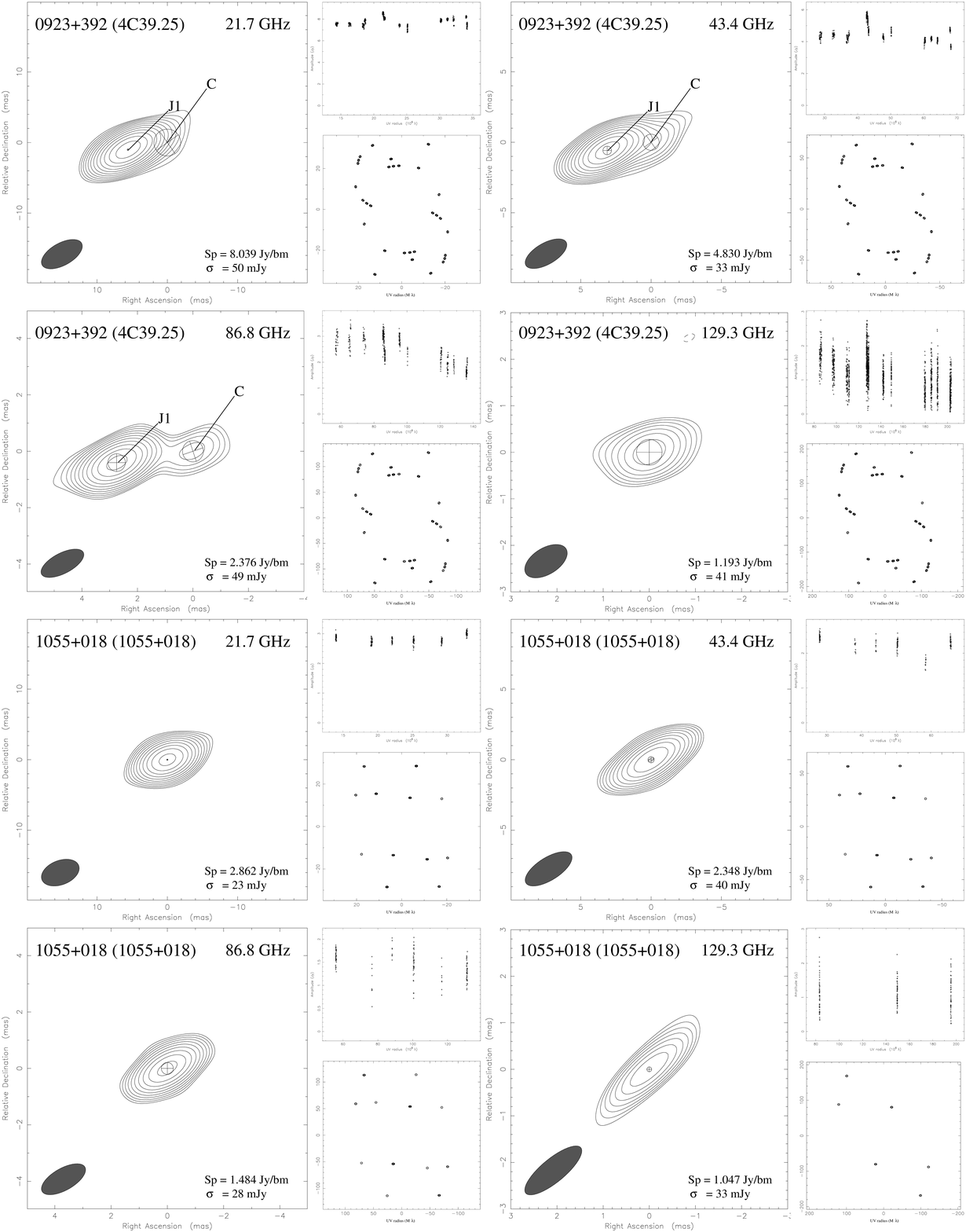}
\caption{{\it CLEAN images (continued)}}
\end{figure*}
\setcounter{figure}{1}
\begin{figure*}[!t]
\epsscale{2}
\plotone{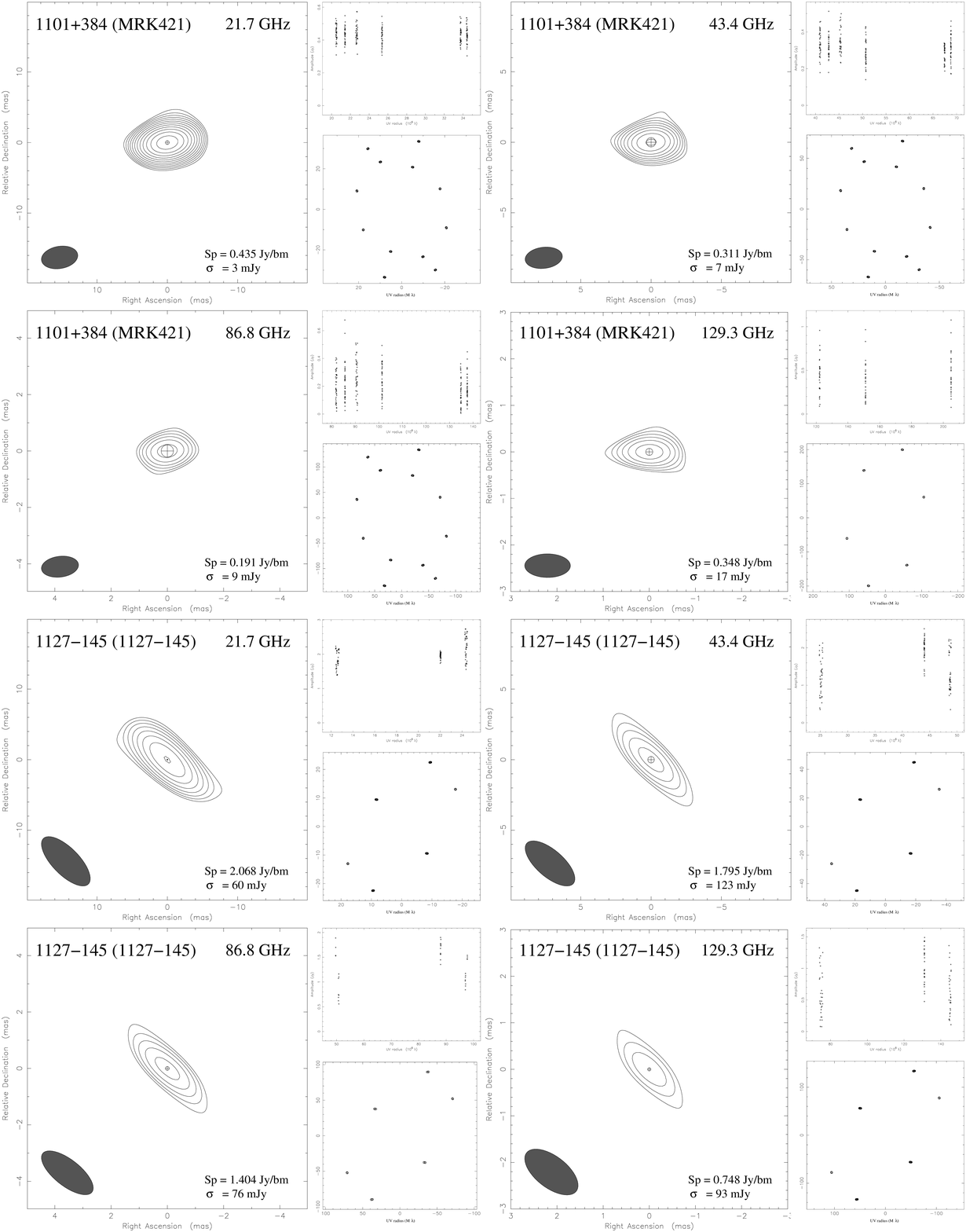}
\caption{{\it CLEAN images (continued)}}
\end{figure*}
\setcounter{figure}{1}
\begin{figure*}[!t]
\epsscale{2}
\plotone{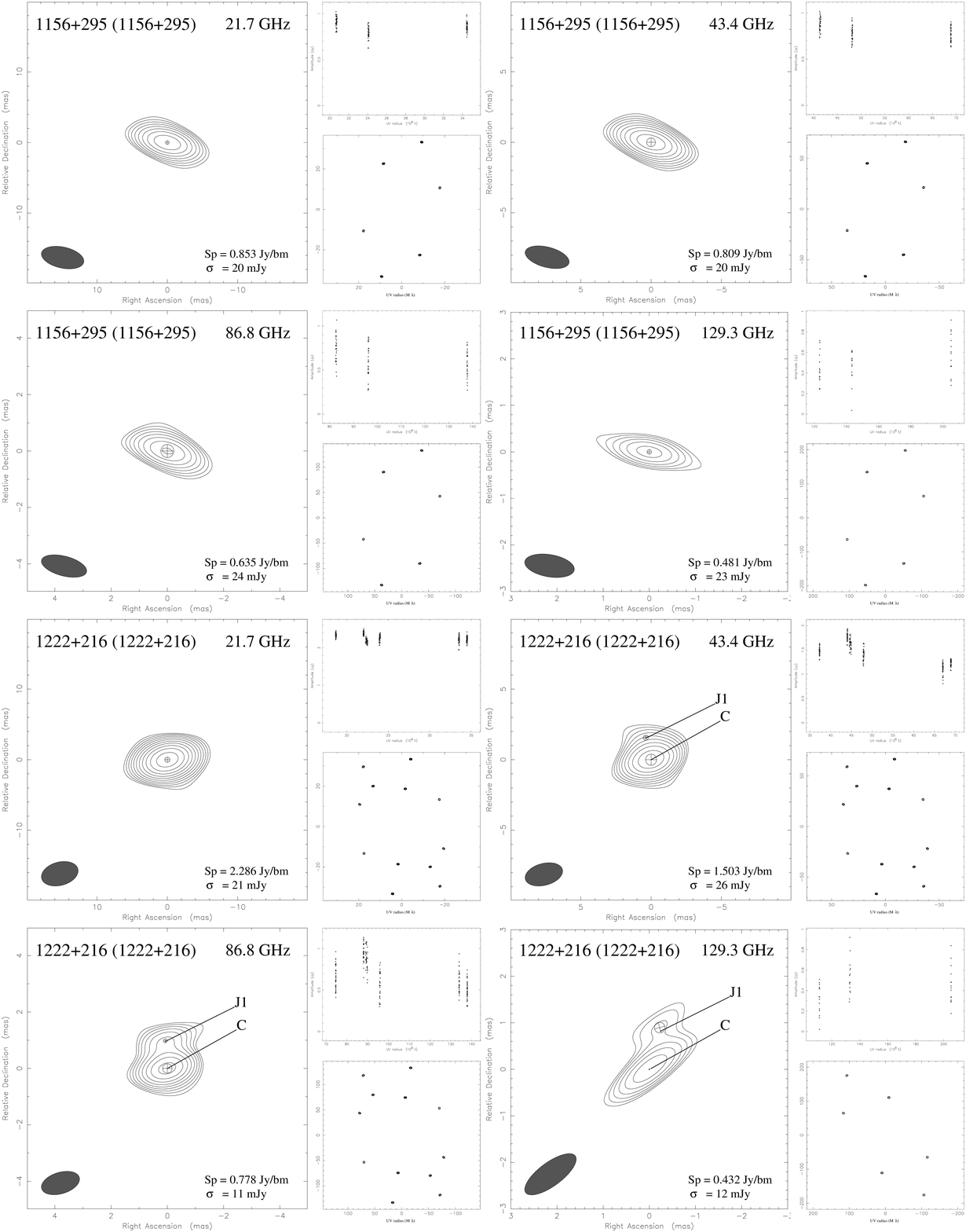}
\caption{{\it CLEAN images (continued)}}
\end{figure*}
\setcounter{figure}{1}
\begin{figure*}[!t]
\epsscale{2}
\plotone{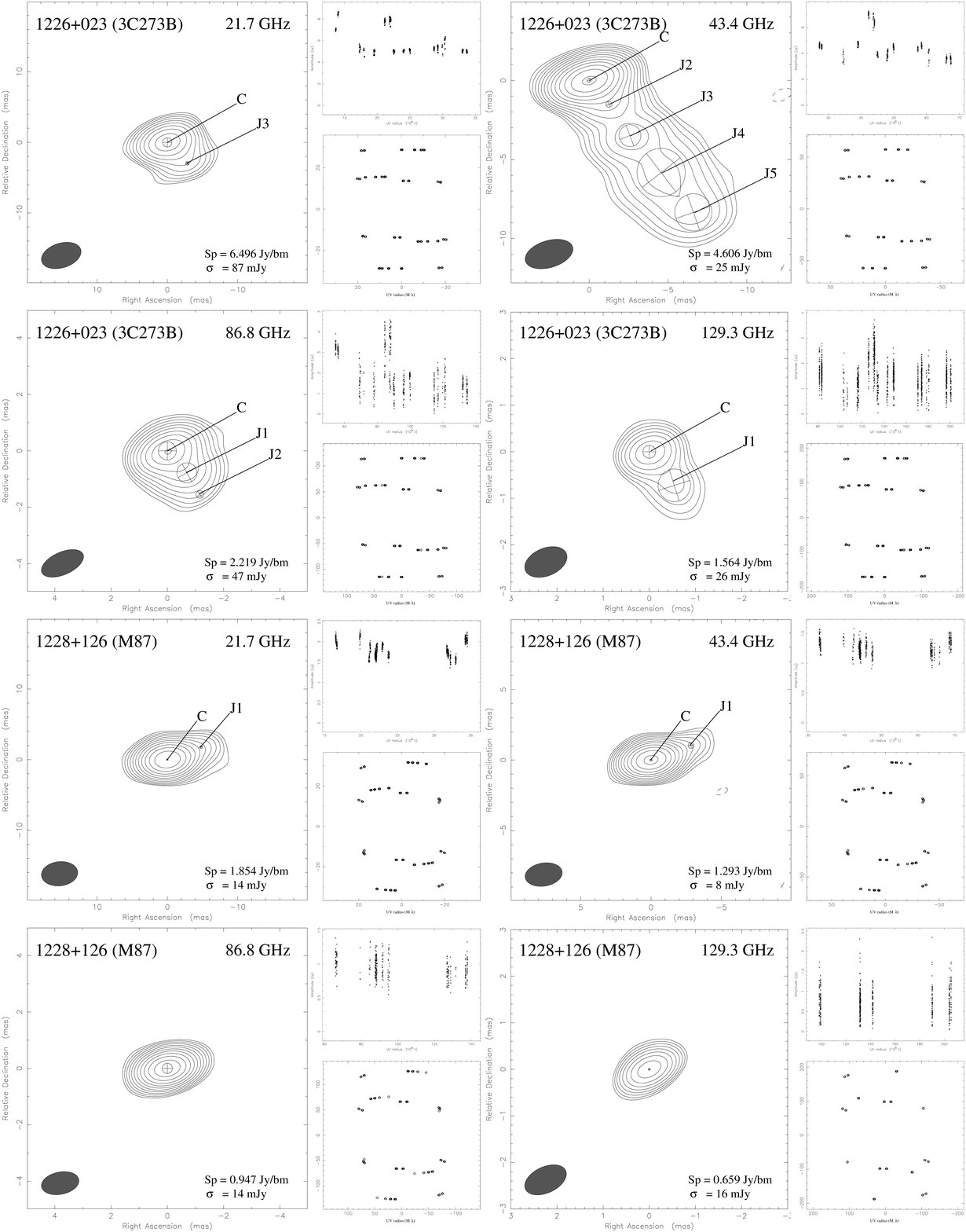}
\caption{{\it CLEAN images (continued)}}
\end{figure*}
\setcounter{figure}{1}
\begin{figure*}[!t]
\epsscale{2}
\plotone{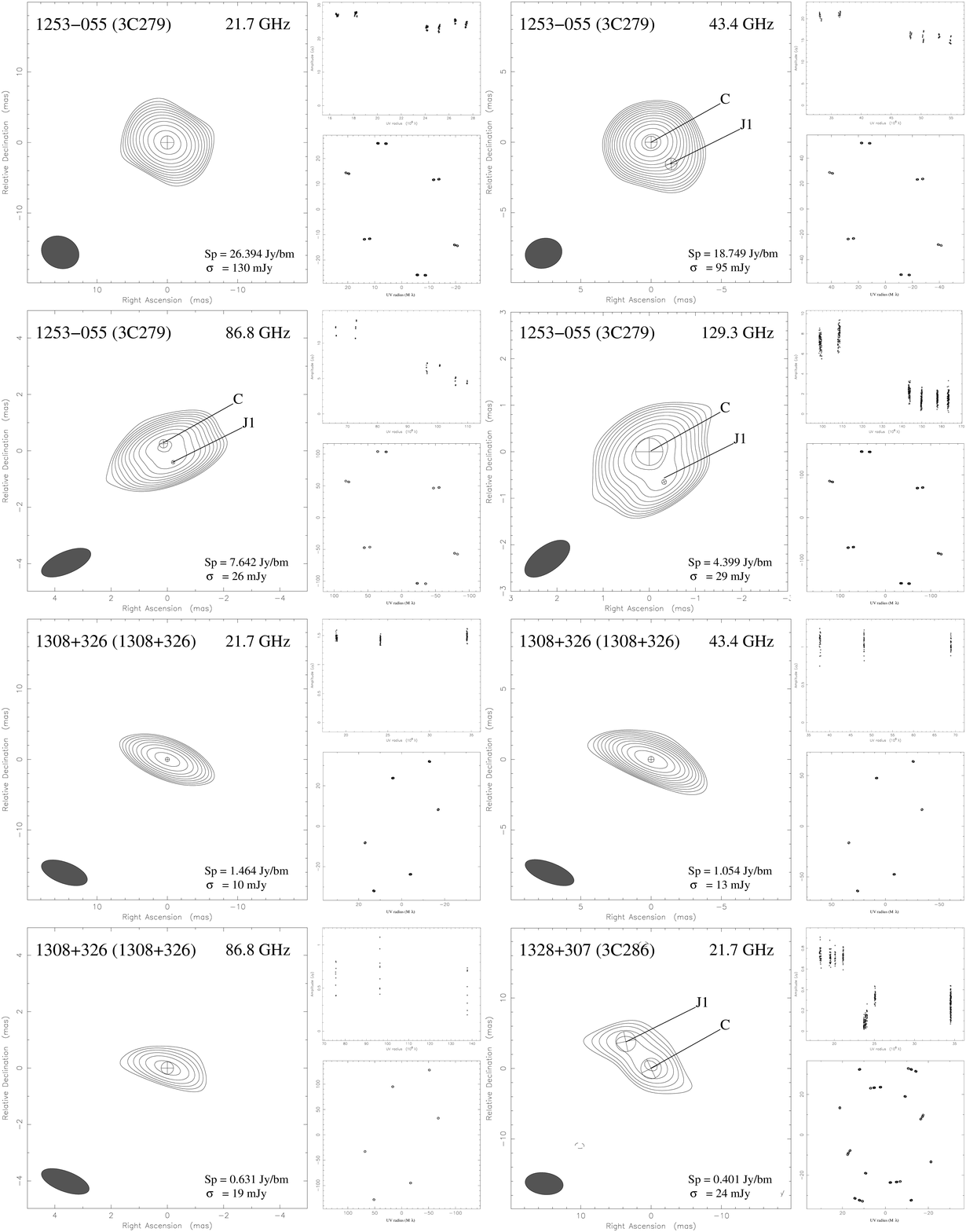}
\caption{{\it CLEAN images (continued)}}
\end{figure*}
\setcounter{figure}{1}
\begin{figure*}[!t]
\epsscale{2}
\plotone{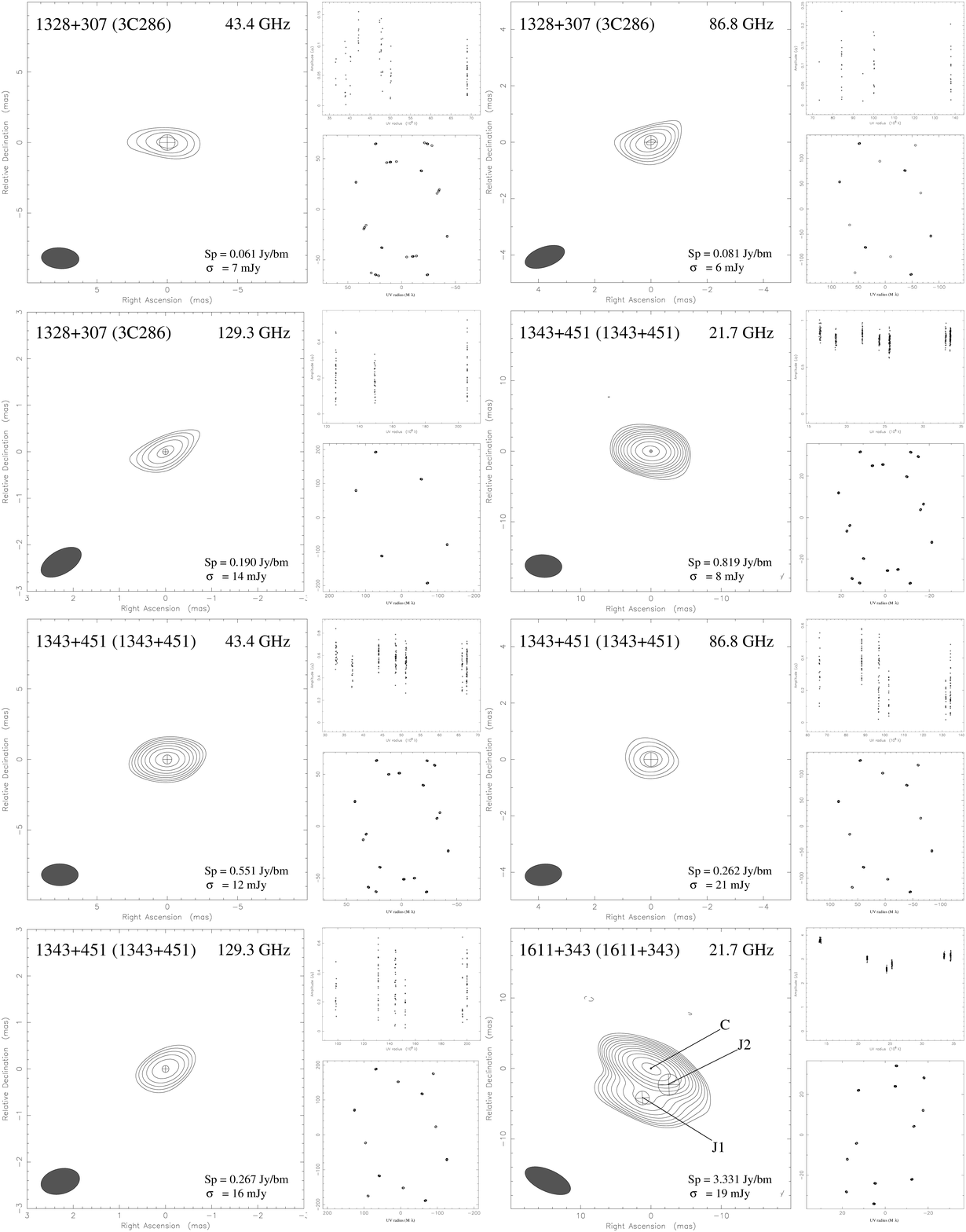}
\caption{{\it CLEAN images (continued)}}
\end{figure*}
\setcounter{figure}{1}
\begin{figure*}[!t]
\epsscale{2}
\plotone{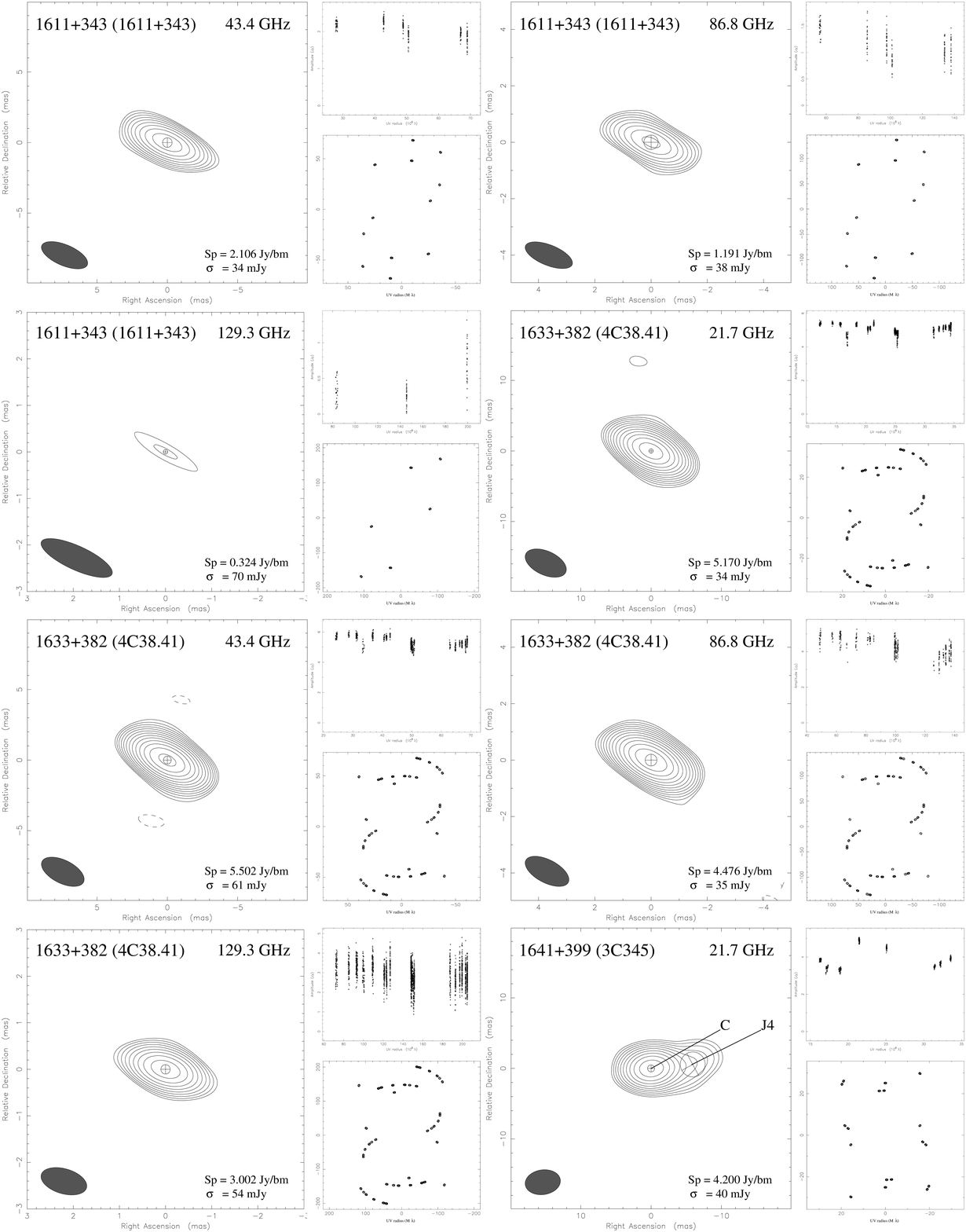}
\caption{{\it CLEAN images (continued)}}
\end{figure*}
\setcounter{figure}{1}
\begin{figure*}[!t]
\epsscale{2}
\plotone{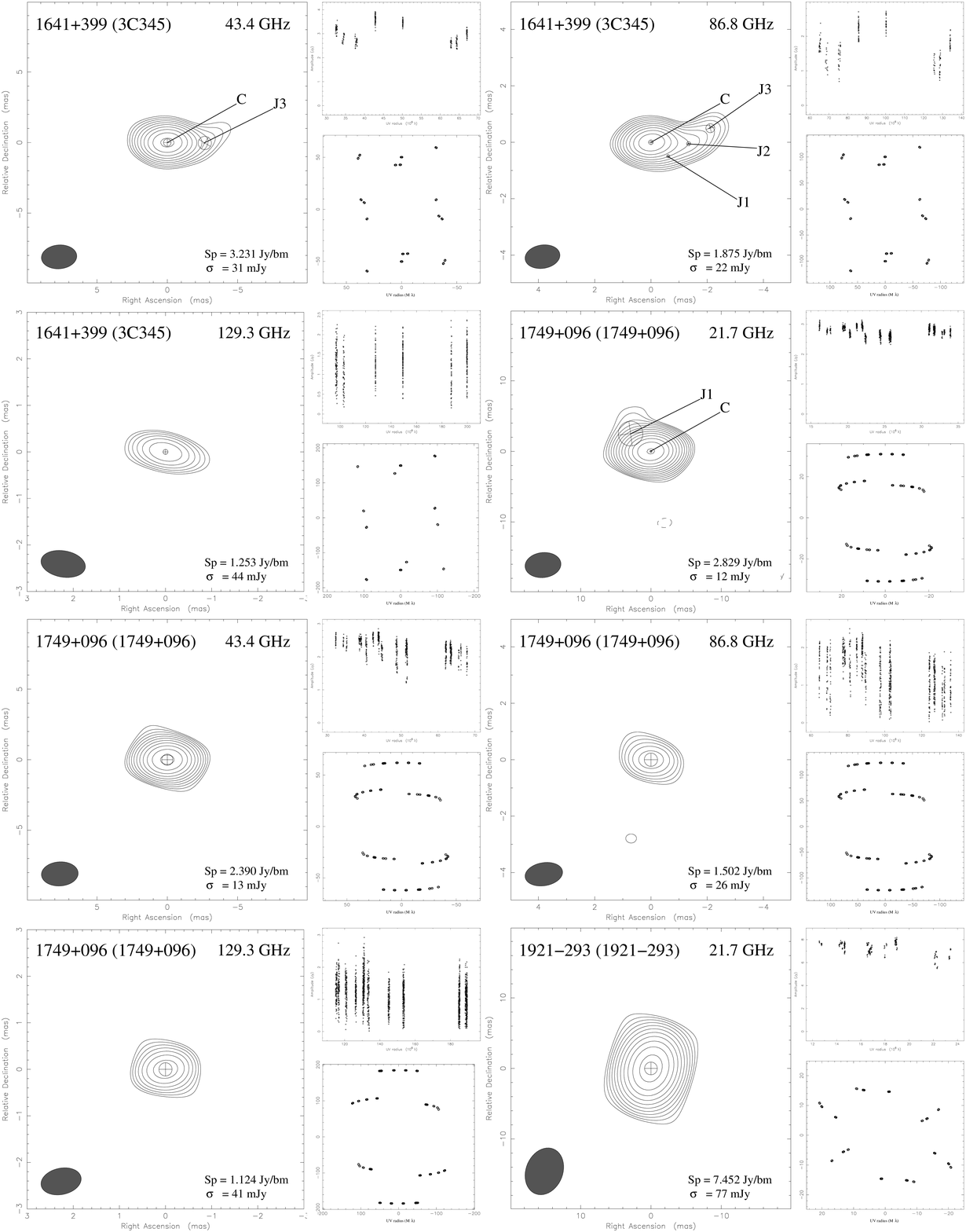}
\caption{{\it CLEAN images (continued)}}
\end{figure*}
\setcounter{figure}{1}
\begin{figure*}[!t]
\epsscale{2}
\plotone{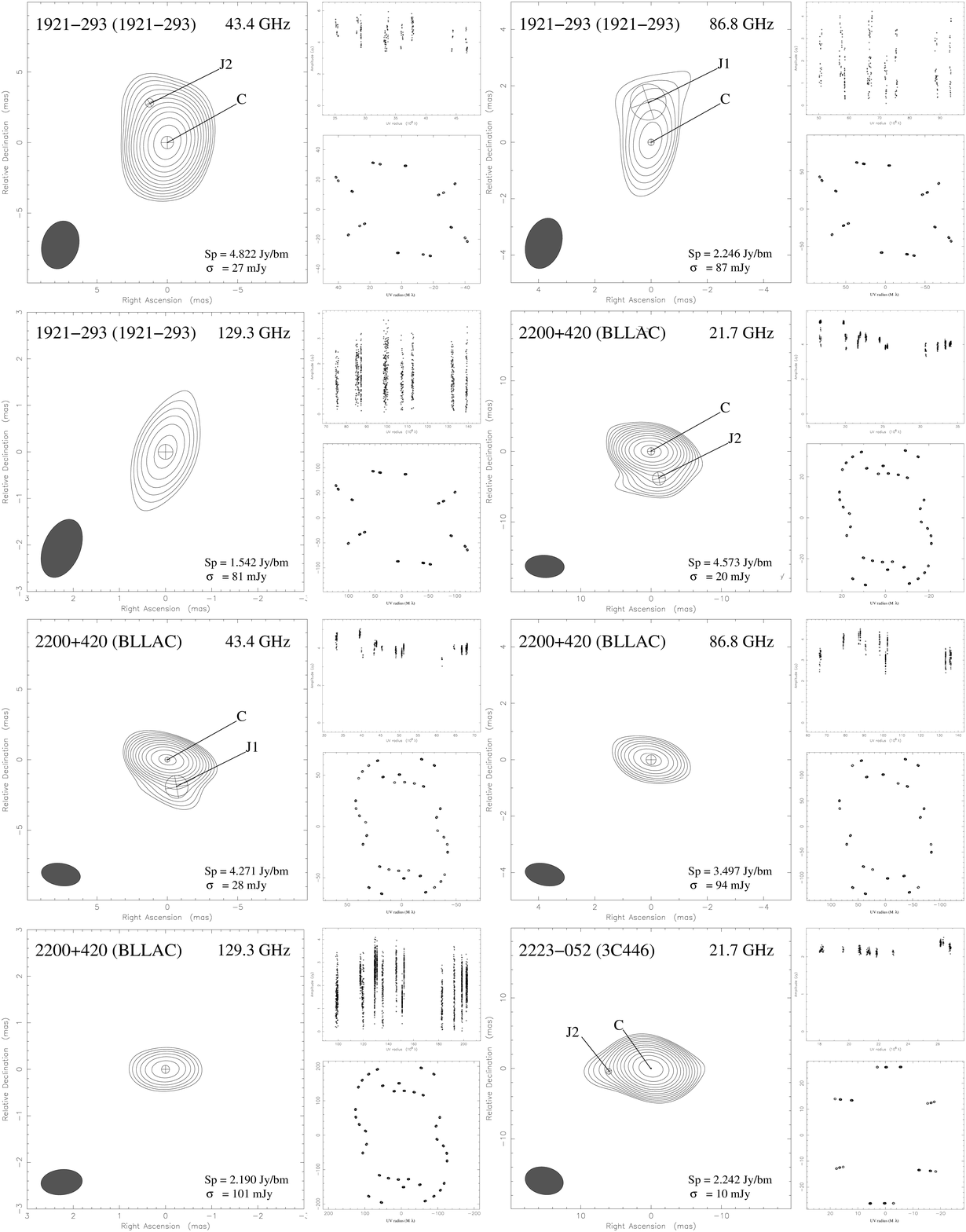}
\caption{{\it CLEAN images (continued)}}
\end{figure*}
\setcounter{figure}{1}
\begin{figure*}[!t]
\epsscale{2}
\plotone{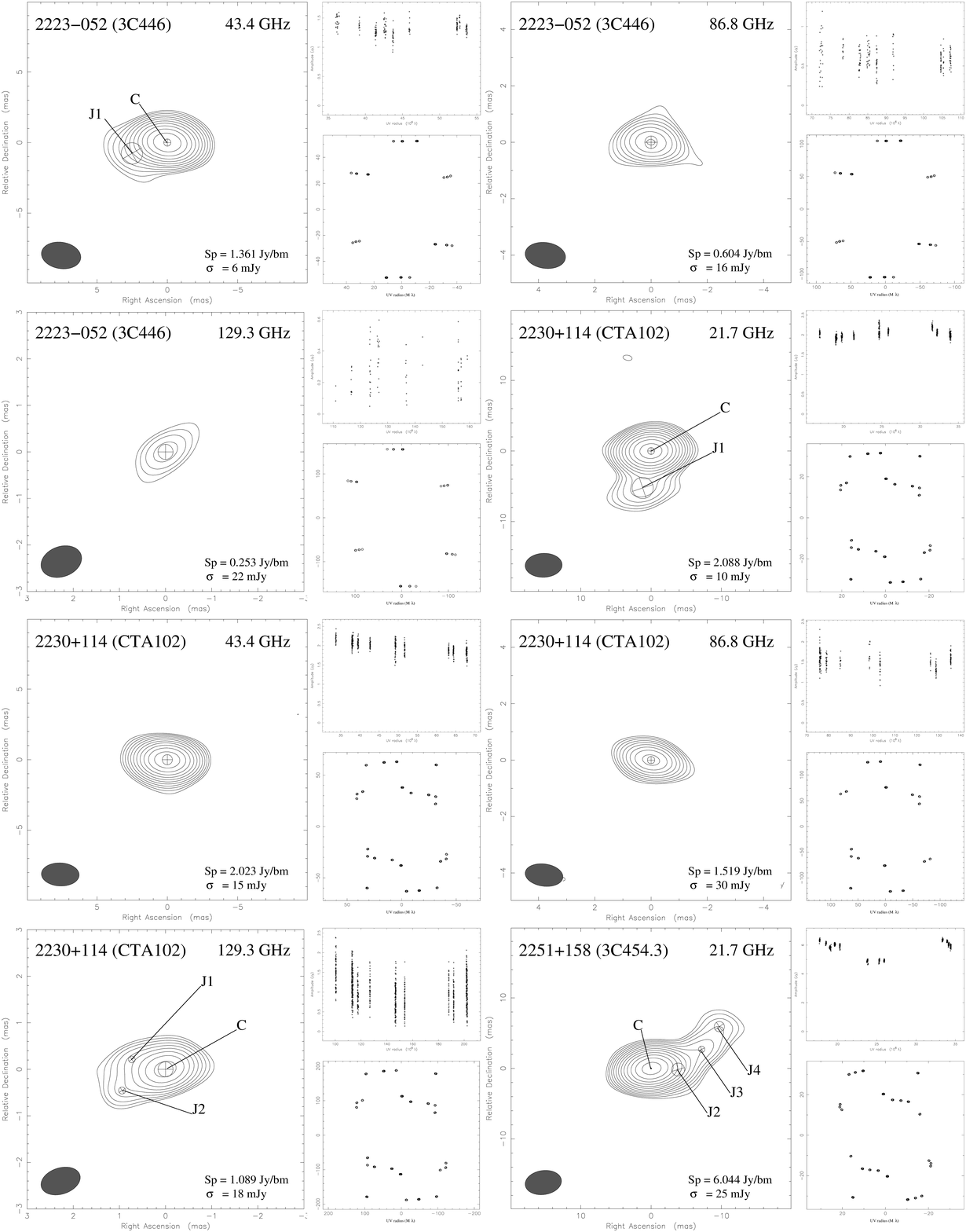}
\caption{{\it CLEAN images (continued)}}
\end{figure*}
\setcounter{figure}{1}
\begin{figure*}[!t]
\epsscale{2}
\plotone{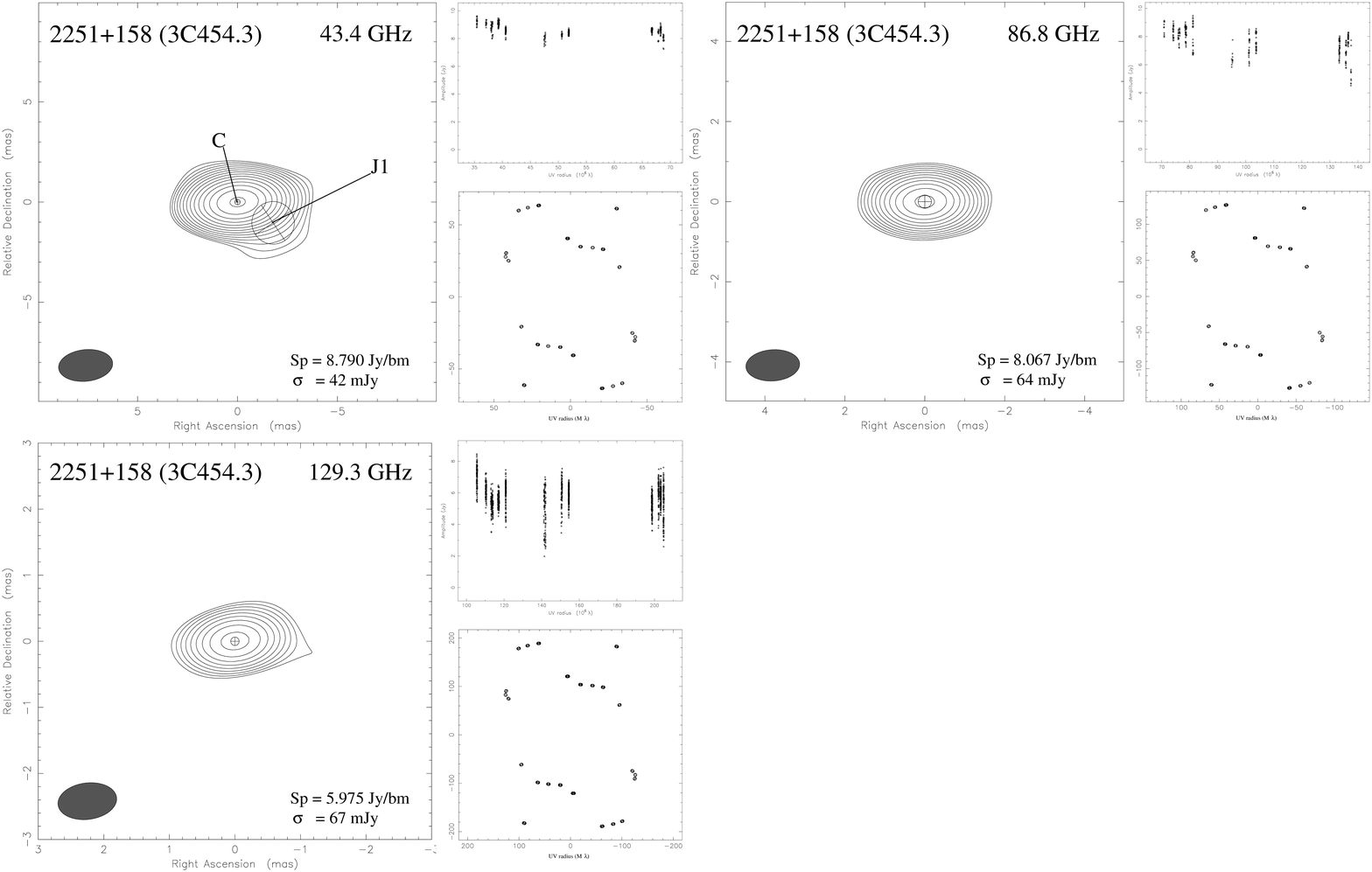}
\caption{{\it CLEAN images (continued)}}
\end{figure*}

\clearpage

\begin{figure*}[!tf]
\epsscale{0.9}
\plotone{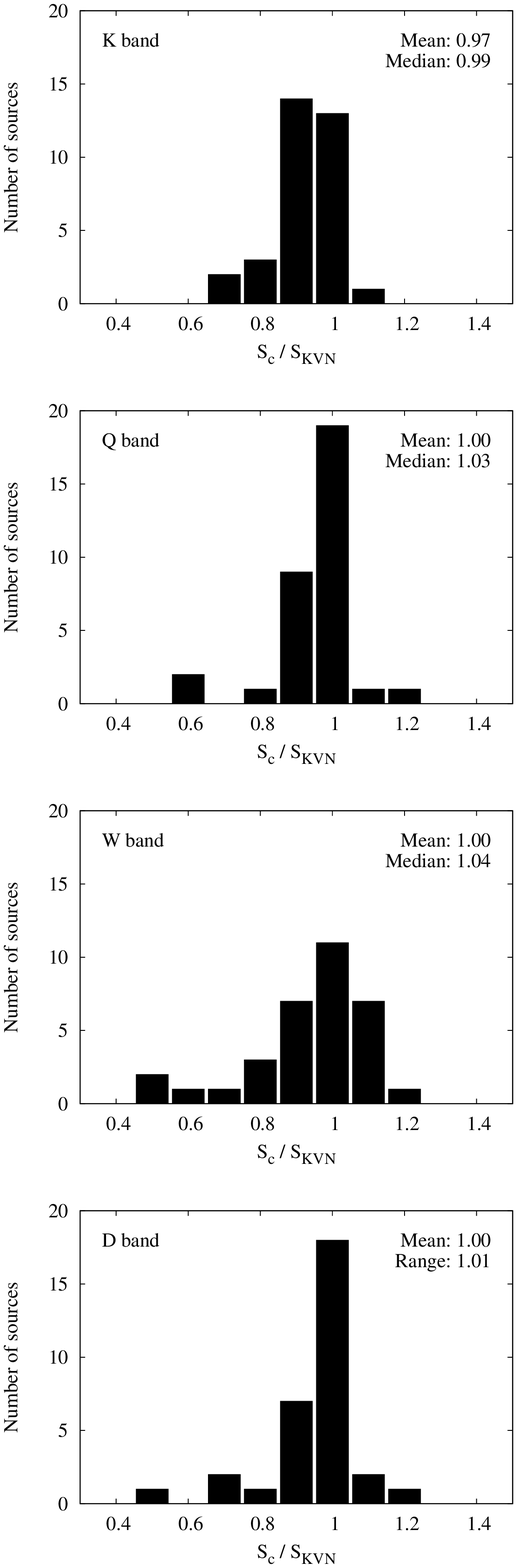}
\caption{Distributions of the ratio of the core flux density ($S_{\rm c}$)
to the CLEAN flux density ($S_{\rm KVN}$).
The mean and median of the distributions are presented.
\label{fig-hist-ratio}}
\end{figure*}


\begin{figure*}[!tf]
\epsscale{0.9}
\plotone{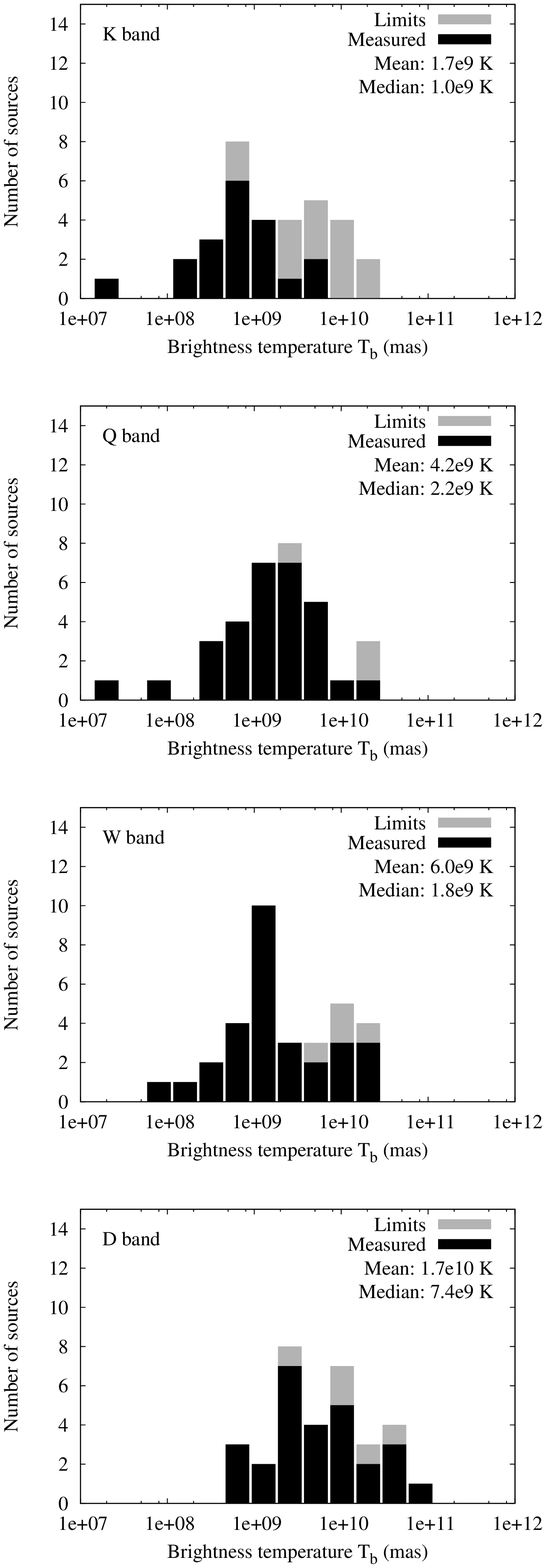}
\caption{Distributions of the brightness temperature of the core components for the imaged sources.
The mean and median of the distributions for the measured sizes are presented.
\label{fig-hist-tb}}
\end{figure*}

\begin{figure*}[!tf]
\epsscale{0.9}
\plotone{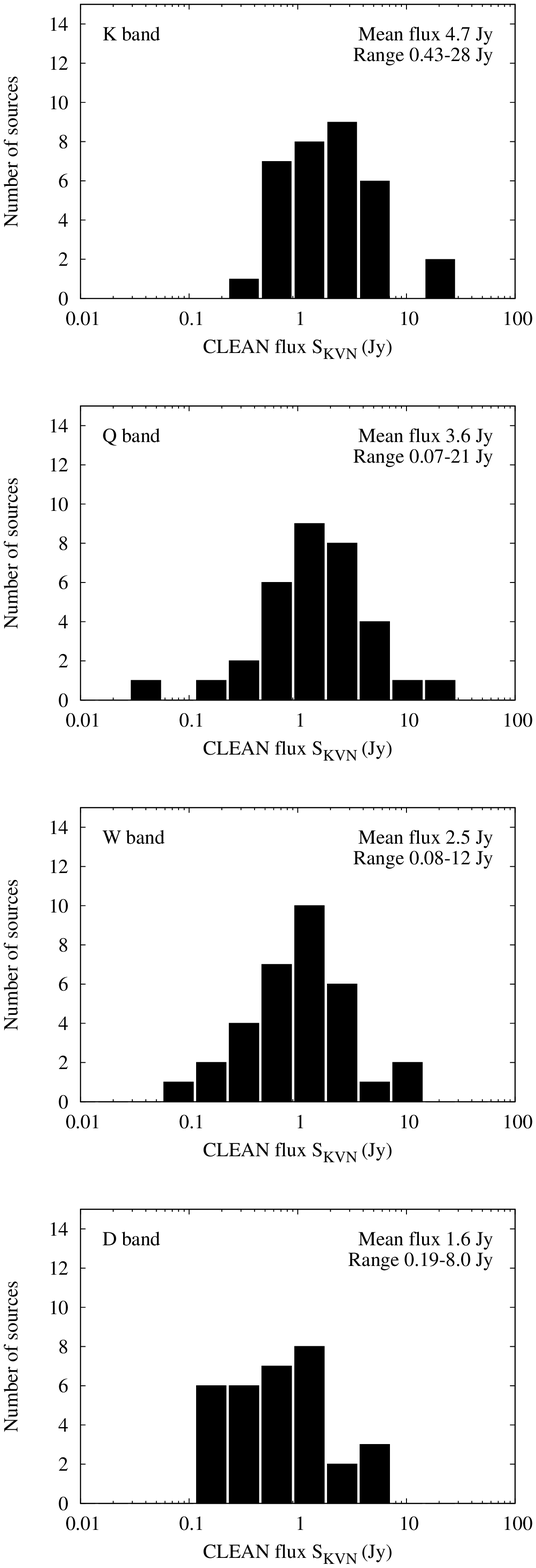}
\caption{Distributions of the CLEAN flux density.
The mean and range of the distributions are presented.
\label{fig-hist-CLEAN}}
\end{figure*}

\begin{figure*}[!t]
\epsscale{1.9}
\plotone{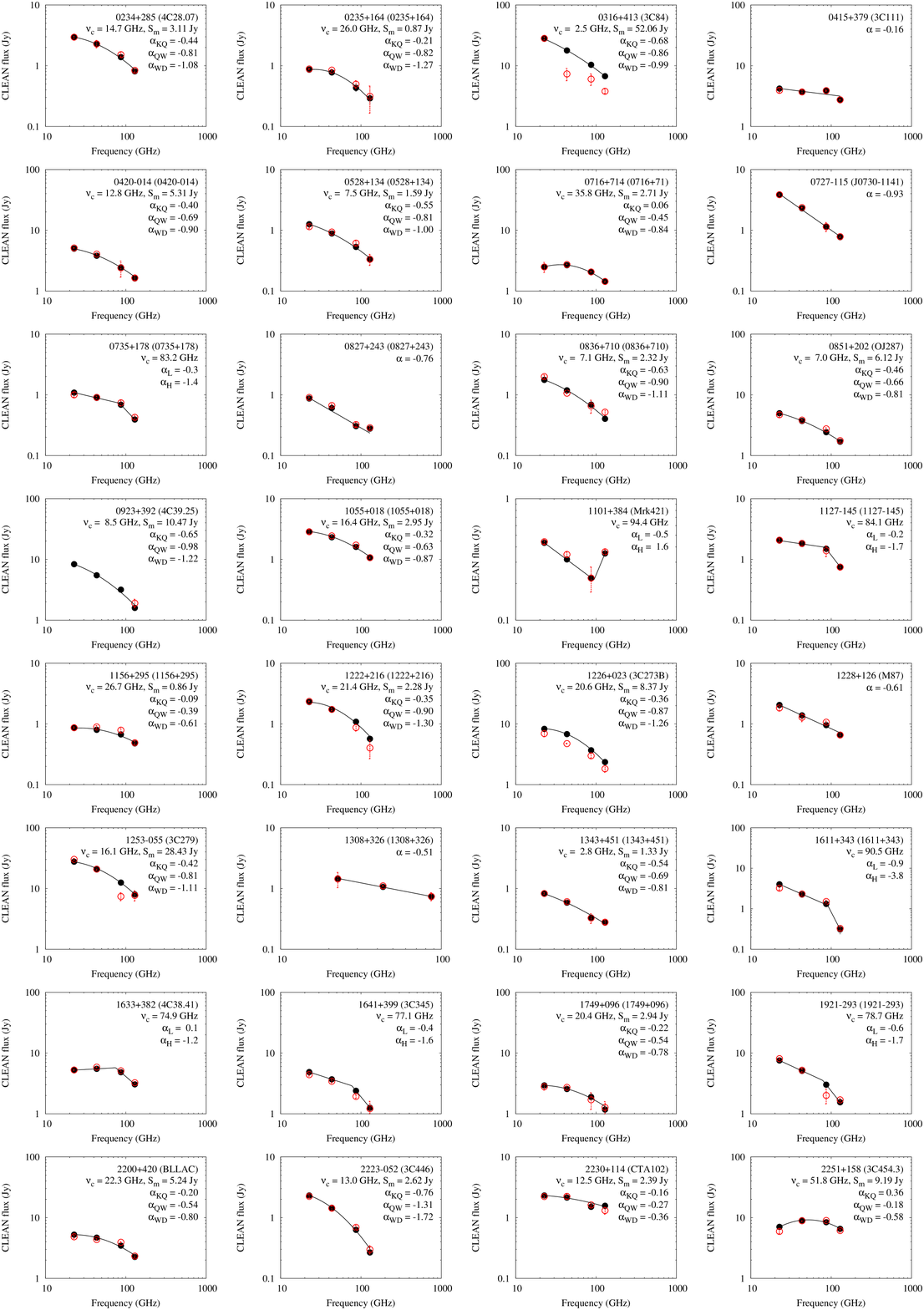}
\caption{Spectra of CLEAN flux density (black dot)
	and core flux density (red circle) for 32 sources imaged at 22-129~GHz.
	Black solid lines are the best fitting power law.
\label{fig-spind}}
\end{figure*}

\begin{figure*}[!tf]
\epsscale{0.9}
\plotone{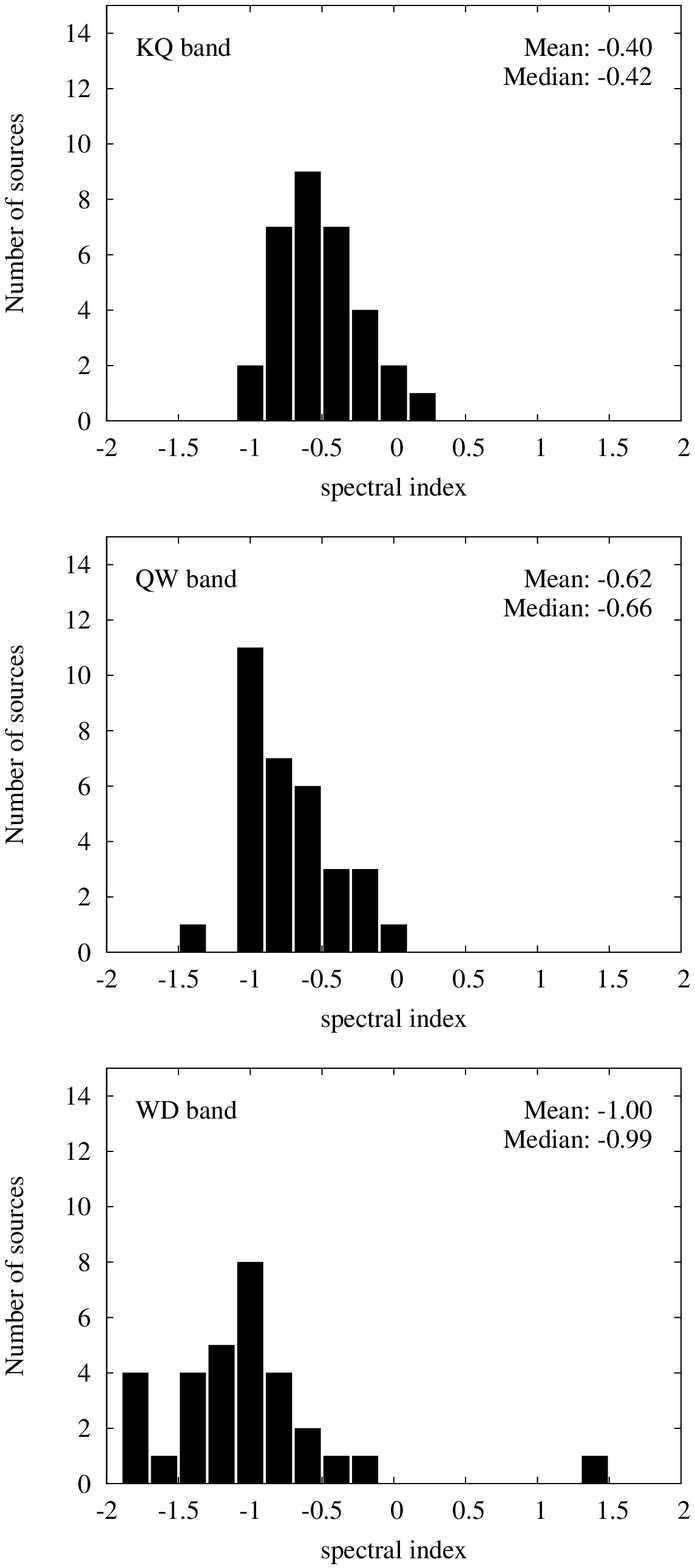}
\caption{Distributions of the spectral indices.
The mean and median of the distributions are presented.
\label{fig-hist-spind}}
\end{figure*}



\begin{figure*}[!tf]
\epsscale{2.0}
\plotone{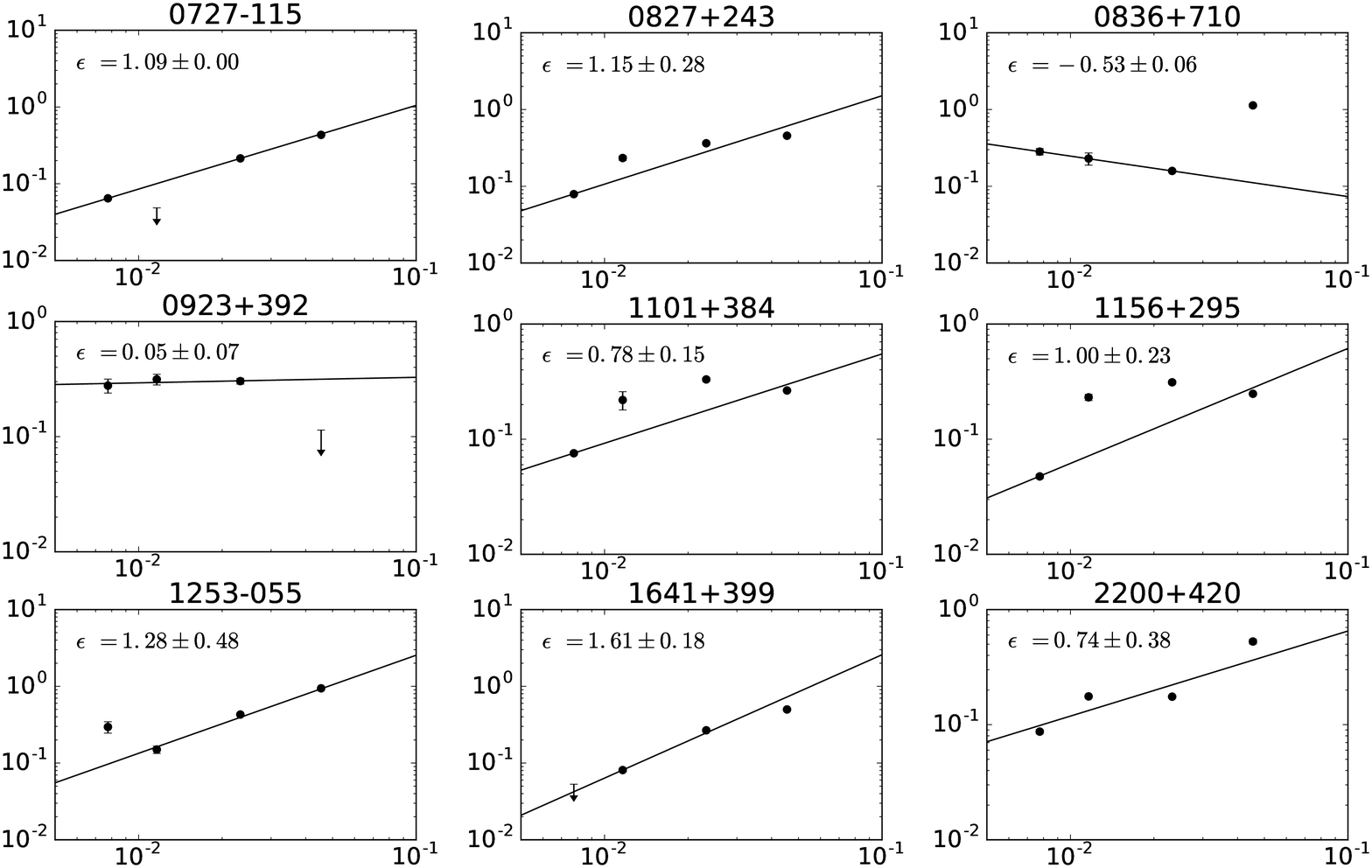}
\caption{Representative subset of plots for the core size (in mas) versus 1/frequency (in 1/GHz).
The best fit to an exponential function is marked with a solid line,
and the value of the exponent is shown in the top left corner.
Left: all points are well aligned with increasing size with distance;
Middle: a convex trend seems apparent;
Right: other patterns.
\label{fig-geometry}}
\end{figure*}


\clearpage

\begin{deluxetable}{lcccccc}
\tabletypesize{\scriptsize}
\tablecaption{Source list\label{t1}}
\tablewidth{0pt}
\tablehead{
\colhead{Source} & 
\colhead{Name} & 
\colhead{Type} &
\colhead{$z$} & 
\colhead{$\alpha_{\rm 2000}$} &
\colhead{$\delta_{\rm 2000}$} &
\colhead{$S_{\rm Fermi}$}\\ 
\colhead{(1)} & 
\colhead{(2)} & 
\colhead{(3)} &
\colhead{(4)} & 
\colhead{(5)} & 
\colhead{(6)} &
\colhead{(7)}
}
\startdata
0218$+$357  &            &  Q   &   0.944 &     02 21 05.468  &  $+$35 56 13.73456  & 4.6e-09 \\
0234$+$285  &   4C +28.07&  Q   &   1.206 &     02 37 52.405  &  $+$28 48 08.98998  & 9.7e-09 \\
0235$+$164  &            &  B   &   0.940 &     02 38 38.930  &  $+$16 36 59.27450  & 1.0e-08 \\
0316$+$413  &       3C 84&  G   &   0.018 &     03 19 48.160  &  $+$41 30 42.10558  & 2.1e-08 \\
0415$+$379  &      3C 111&  G   &   0.049 &     04 18 21.277  &  $+$38 01 35.80018  & ...   \\
0420$-$014  &            &  Q   &   0.916 &     04 23 15.800  &  $-$01 20 33.06558  & 5.6e-09 \\
0528$+$134  &            &  Q   &   2.070 &     05 30 56.416  &  $+$13 31 55.14944  & 3.7e-09 \\
0716$+$714  &            &  B   &   0.127 &     07 21 53.448  &  $+$71 20 36.36340  & 2.2e-08 \\
0727$-$115  &            &  Q   &   1.591 &     07 30 19.112  &  $-$11 41 12.60064  & 1.8e-08 \\
0735$+$178  &            &  B   &   0.450 &     07 38 07.393  &  $+$17 42 18.99811  & ...   \\
0827$+$243  &            &  Q   &   0.942 &     08 30 52.086  &  $+$24 10 59.82027  & 1.1e-09 \\
0836$+$710  &            &  Q   &   2.218 &     08 41 24.365  &  $+$70 53 42.17302  & 1.3e-09 \\
0851$+$202  &      OJ 287&  B   &   0.306 &     08 54 48.874  &  $+$20 06 30.64077  & 5.9e-09 \\
0923$+$392  &   4C +39.25&  Q   &   0.695 &     09 27 03.013  &  $+$39 02 20.85177  & ...   \\
1055$+$018  &            &  Q   &   0.888 &     10 58 29.605  &  $+$01 33 58.82359  & 6.5e-09 \\
1101$+$384  &     Mrk 421&  B   &   0.031 &     11 04 27.313  &  $+$38 12 31.79895  & 3.0e-08 \\
1127$-$145  &            &  Q   &   1.184 &     11 30 07.052  &  $-$14 49 27.38835  & 9.0e-10 \\
1156$+$295  &            &  Q   &   0.725 &     11 59 31.833  &  $+$29 14 43.82678  & 8.4e-09 \\
1222$+$216  &            &  Q   &   0.434 &     12 24 54.458  &  $+$21 22 46.38856  & 2.6e-08 \\
1226$+$023  &     3C 273B&  Q   &   0.158 &     12 29 06.699  &  $+$02 03 08.59796  & 9.4e-09 \\
1228$+$126  &         M87&  G   &   0.004 &     12 30 49.423  &  $+$12 23 28.04365  & 1.3e-09 \\
1253$-$055  &      3C 279&  Q   &   0.536 &     12 56 11.166  &  $-$05 47 21.52490  & 2.1e-08 \\
1308$+$326  &            &  Q   &   0.997 &     13 10 28.663  &  $+$32 20 43.78277  & 3.7e-09 \\
1328$+$307  &      3C 286&  Q   &   0.846 &     13 31 08.288  &  $+$30 30 32.95924  & ...   \\
1343$+$451  &            &  Q   &   2.534 &     13 45 33.172  &  $+$44 52 59.57257  & 6.5e-09 \\
1611$+$343  &            &  Q   &   1.400 &     16 13 41.064  &  $+$34 12 47.90878  & 6.5e-10 \\
1633$+$382  &   4C +38.41&  Q   &   1.813 &     16 35 15.492  &  $+$38 08 04.50043  & 1.1e-08 \\
1641$+$399  &      3C 345&  Q   &   0.593 &     16 42 58.809  &  $+$39 48 36.99385  & 2.5e-09 \\
1749$+$096  &            &  B   &   0.322 &     17 51 32.818  &  $+$09 39 00.72829  & 4.3e-09 \\
1921$-$293  &            &  Q   &   0.353 &     19 24 51.055  &  $-$29 14 30.12115  & 2.2e-09 \\
2200$+$420  &            &  B   &   0.069 &     22 02 43.291  &  $+$42 16 39.97986  & 1.6e-08 \\
2223$-$052  &      3C 446&  Q   &   1.404 &     22 25 47.259  &  $-$04 57 01.39090  & 1.9e-09 \\
2230$+$114  &     CTA 102&  Q   &   1.037 &     22 32 36.408  &  $+$11 43 50.90394  & 5.0e-09 \\
2251$+$158  &    3C 454.3&  Q   &   0.859 &     22 53 57.747  &  $+$16 08 53.56073  & 1.1e-07 \\
\enddata
\tablecomments{ 
Column designation: 1~-~IAU source name; 2~-~common name;
3~-~optical type obtained from \cite{vv06};
4~-~redshift; 
5~-~Right Ascension (J2000) [hh mm ss];
6~-~declination (J2000) [dd mm ss];
7~-~gamma-ray flux density (ph~cm$^{-2}$~s$^{-1}$). 
}
\end{deluxetable}

\begin{deluxetable}{lcccrrrrrr}
\tabletypesize{\scriptsize}
\tablecaption{Image parameters\label{t2}}
\tablewidth{0pt}
\tablehead{
\colhead{Name} & 
\colhead{Band} & 
\colhead{$B_{\rm maj}$} &
\colhead{$B_{\rm min}$} & 
\colhead{$B_{\rm PA}$} & 
\colhead{$S_{\rm KVN}$} &
\colhead{$S_{\rm p}$} &
\colhead{$\sigma$} &
\colhead{$D$} &
\colhead{$\xi_{\rm r}$}\\ 
\colhead{(1)} & 
\colhead{(2)} & 
\colhead{(3)} &
\colhead{(4)} & 
\colhead{(5)} & 
\colhead{(6)} &
\colhead{(7)} &
\colhead{(8)} &
\colhead{(9)} &
\colhead{(10)}
}
\startdata
0234$+$285 &K   &      5.533 &      3.076 &       83.8 &       2.957 &       2.824 &          11 &    260 &  0.68 \\
           &Q   &      2.906 &      1.463 &       81.8 &       2.262 &       2.076 &          11 &    195 &  0.65 \\
           &W   &      1.459 &      0.723 &       82.9 &       1.384 &       1.234 &          35 &     35 &  0.70 \\
           &D   &      0.959 &      0.535 &  $-$86.1 &       0.808 &       0.784 &          38 &     21 &  0.63 \\
0235$+$164 &K   &      7.635 &      3.025 &  $-$57.2 &       0.862 &       0.857 &          12 &     72 &  0.54 \\
           &Q   &      3.833 &      1.458 &  $-$60.0 &       0.775 &       0.755 &          17 &     44 &  0.62 \\
           &W   &      1.887 &      0.701 &  $-$60.4 &       0.432 &       0.420 &          54 &      8 &  0.49 \\
           &D   &      1.495 &      0.509 &  $-$53.4 &       0.290 &       0.275 &          46 &      6 &  0.50 \\
0316$+$413 &K   &      5.866 &      2.790 &       88.3 &      27.795 &      23.617 &         162 &    146 &  0.62 \\
           &Q   &      3.075 &      1.420 &  $-$82.3 &      17.830 &      11.196 &          56 &    201 &  0.77 \\
           &W   &      1.446 &      0.688 &       80.9 &      10.350 &       4.951 &          67 &     73 &  0.71 \\
           &D   &      0.939 &      0.517 &  $-$83.5 &       6.720 &       3.250 &          19 &    174 &  0.72 \\
0415$+$379 &K   &      6.363 &      3.020 &       72.0 &       4.204 &       3.953 &          24 &    163 &  0.61 \\
           &Q   &      3.379 &      1.435 &       69.7 &       3.714 &       3.708 &          24 &    152 &  0.69 \\
           &W   &      1.710 &      0.712 &       69.7 &       3.830 &       3.724 &          30 &    124 &  0.59 \\
           &D   &      1.096 &      0.498 &       76.8 &       2.706 &       2.688 &          36 &     75 &  0.52 \\
0420$-$014 &K   &      6.414 &      3.265 &  $-$63.3 &       4.984 &       5.058 &          81 &     62 &  0.52 \\
           &Q   &      3.297 &      1.604 &  $-$65.1 &       3.814 &       3.777 &          50 &     76 &  0.46 \\
           &W   &      1.664 &      0.800 &  $-$66.3 &       2.452 &       2.453 &          17 &    144 &  0.46 \\
           &D   &      1.085 &      0.553 &  $-$52.2 &       1.610 &       1.616 &          56 &     29 &  0.44 \\
0528$+$134 &K   &      5.360 &      3.323 &       86.2 &       1.259 &       1.159 &           6 &    199 &  0.53 \\
           &Q   &      2.767 &      1.566 &       82.3 &       0.887 &       0.878 &          11 &     82 &  0.55 \\
           &W   &      1.344 &      0.776 &       86.7 &       0.532 &       0.519 &          19 &     27 &  0.56 \\
           &D   &      0.888 &      0.566 &  $-$81.5 &       0.337 &       0.335 &          18 &     19 &  0.55 \\
0716$+$714 &K   &      5.826 &      4.243 &  $-$61.5 &       2.507 &       2.511 &          17 &    145 &  0.58 \\
           &Q   &      2.897 &      1.992 &  $-$69.9 &       2.658 &       2.654 &          21 &    127 &  0.61 \\
           &W   &      1.462 &      0.975 &  $-$70.8 &       2.064 &       2.045 &          31 &     66 &  0.58 \\
           &D   &      0.970 &      0.681 &  $-$60.8 &       1.431 &       1.424 &          20 &     73 &  0.61 \\
0727$-$115 &K   &      6.929 &      3.570 &  $-$54.7 &       3.821 &       3.754 &          23 &    166 &  0.48 \\
           &Q   &      3.589 &      1.749 &  $-$58.2 &       2.298 &       2.291 &          13 &    176 &  0.44 \\
           &W   &      1.757 &      0.888 &  $-$57.4 &       1.155 &       1.155 &          12 &     95 &  0.43 \\
           &D   &      1.206 &      0.588 &  $-$45.9 &       0.770 &       0.766 &           5 &    158 &  0.48 \\
0735$+$178 &K   &      7.012 &      3.033 &  $-$59.8 &       1.088 &       1.013 &           4 &    250 &  0.62 \\
           &Q   &      3.521 &      1.456 &  $-$62.7 &       0.891 &       0.836 &           7 &    127 &  0.60 \\
           &W   &      1.867 &      0.708 &  $-$61.1 &       0.687 &       0.569 &          18 &     31 &  0.59 \\
           &D   &      1.050 &      0.501 &  $-$57.3 &       0.392 &       0.366 &          14 &     27 &  0.48 \\
0827$+$243 &K   &      9.665 &      3.073 &       60.9 &       0.872 &       0.868 &          19 &     46 &  0.46 \\
           &Q   &      5.178 &      1.436 &       60.8 &       0.606 &       0.592 &          20 &     29 &  0.48 \\
           &W   &      2.603 &      0.713 &       61.2 &       0.305 &       0.273 &           7 &     38 &  0.47 \\
           &D   &      1.614 &      0.521 &       64.0 &       0.278 &       0.275 &          18 &     15 &  0.47 \\
0836$+$710 &K   &      5.680 &      4.162 &  $-$53.6 &       1.762 &       1.648 &          11 &    148 &  0.55 \\
           &Q   &      2.924 &      1.998 &  $-$52.8 &       1.187 &       1.048 &           4 &    264 &  0.64 \\
           &W   &      1.466 &      0.991 &  $-$53.7 &       0.688 &       0.601 &          16 &     38 &  0.55 \\
           &D   &      0.926 &      0.700 &  $-$50.8 &       0.403 &       0.336 &          10 &     33 &  0.58 \\
0851$+$202 &K   &      6.126 &      3.080 &  $-$65.4 &       4.974 &       4.746 &          15 &    317 &  0.74 \\
           &Q   &      3.223 &      1.480 &  $-$66.3 &       3.716 &       3.683 &          25 &    149 &  0.64 \\
           &W   &      1.585 &      0.734 &  $-$64.7 &       2.425 &       2.453 &          48 &     51 &  0.70 \\
           &D   &      1.067 &      0.512 &  $-$58.5 &       1.692 &       1.651 &          47 &     35 &  0.59 \\
0923$+$392 &K   &      6.429 &      3.181 &  $-$61.0 &       8.338 &       8.039 &          50 &    162 &  0.65 \\
           &Q   &      3.314 &      1.515 &  $-$60.1 &       5.475 &       4.830 &          33 &    147 &  0.62 \\
           &W   &      1.697 &      0.730 &  $-$63.0 &       3.173 &       2.376 &          49 &     48 &  0.59 \\
           &D   &      1.023 &      0.569 &  $-$58.8 &       1.588 &       1.193 &          41 &     29 &  0.67 \\
1055$+$018 &K   &      5.620 &      3.449 &  $-$68.7 &       2.879 &       2.862 &          23 &    123 &  0.53 \\
           &Q   &      3.807 &      1.617 &  $-$58.0 &       2.314 &       2.348 &          40 &     58 &  0.48 \\
           &W   &      1.754 &      0.781 &  $-$60.4 &       1.607 &       1.484 &          28 &     53 &  0.57 \\
           &D   &      1.546 &      0.510 &  $-$50.7 &       1.053 &       1.047 &          33 &     32 &  0.47 \\
1101$+$384 &K   &      5.229 &      3.170 &  $-$80.4 &       0.433 &       0.435 &           3 &    143 &  0.51 \\
           &Q   &      2.663 &      1.508 &  $-$83.3 &       0.316 &       0.311 &           7 &     43 &  0.56 \\
           &W   &      1.329 &      0.741 &  $-$82.8 &       0.222 &       0.191 &           9 &     20 &  0.54 \\
           &D   &      0.992 &      0.509 &       89.8 &       0.354 &       0.348 &          17 &     21 &  0.45 \\
1127$-$145 &K   &      9.106 &      3.832 &       44.7 &       2.040 &       2.068 &          60 &     34 &  0.42 \\
           &Q   &      4.378 &      1.913 &       49.2 &       1.798 &       1.795 &         123 &     15 &  0.49 \\
           &W   &      2.228 &      0.933 &       51.9 &       1.496 &       1.404 &          76 &     19 &  0.52 \\
           &D   &      1.359 &      0.683 &       52.7 &       0.754 &       0.748 &          93 &      8 &  0.49 \\
1156$+$295 &K   &      6.153 &      2.993 &       77.5 &       0.858 &       0.853 &          20 &     44 &  0.43 \\
           &Q   &      3.218 &      1.437 &       75.3 &       0.800 &       0.809 &          20 &     40 &  0.41 \\
           &W   &      1.656 &      0.702 &       75.3 &       0.672 &       0.635 &          24 &     26 &  0.48 \\
           &D   &      1.089 &      0.480 &       80.5 &       0.478 &       0.481 &          23 &     21 &  0.46 \\
1222$+$216 &K   &      5.343 &      3.317 &  $-$75.0 &       2.298 &       2.286 &          21 &    107 &  0.57 \\
           &Q   &      2.728 &      1.553 &  $-$75.3 &       1.744 &       1.503 &          26 &     59 &  0.59 \\
           &W   &      1.399 &      0.760 &  $-$73.3 &       1.086 &       0.778 &          11 &     68 &  0.61 \\
           &D   &      1.336 &      0.506 &  $-$53.7 &       0.569 &       0.432 &          12 &     36 &  0.43 \\
1226$+$023 &K   &      5.816 &      3.493 &  $-$72.7 &       8.297 &       6.496 &          87 &     74 &  0.68 \\
           &Q   &      3.130 &      1.708 &  $-$73.2 &       6.802 &       4.606 &          25 &    187 &  0.72 \\
           &W   &      1.604 &      0.801 &  $-$66.0 &       3.688 &       2.219 &          47 &     47 &  0.69 \\
           &D   &      0.963 &      0.596 &  $-$67.6 &       2.354 &       1.564 &          26 &     61 &  0.71 \\
1228$+$126 &K   &      5.186 &      3.445 &  $-$83.9 &       2.052 &       1.854 &          14 &    129 &  0.60 \\
           &Q   &      2.661 &      1.642 &  $-$80.9 &       1.383 &       1.293 &           8 &    158 &  0.57 \\
           &W   &      1.353 &      0.793 &  $-$78.8 &       0.956 &       0.947 &          14 &     70 &  0.66 \\
           &D   &      0.962 &      0.561 &  $-$65.4 &       0.663 &       0.659 &          16 &     40 &  0.54 \\
1253$-$055 &K   &      5.499 &      4.506 &       66.6 &      27.655 &      26.394 &         130 &    202 &  0.53 \\
           &Q   &      2.659 &      2.116 &  $-$74.7 &      20.980 &      18.749 &          95 &    197 &  0.54 \\
           &W   &      1.885 &      0.777 &  $-$66.9 &      12.517 &       7.642 &          26 &    298 &  0.64 \\
           &D   &      1.141 &      0.551 &  $-$54.4 &       7.983 &       4.399 &          29 &    154 &  0.81 \\
1308$+$326 &K   &      6.875 &      3.004 &       70.0 &       1.464 &       1.464 &          10 &    140 &  0.45 \\
           &Q   &      3.685 &      1.411 &       69.1 &       1.062 &       1.054 &          13 &     84 &  0.43 \\
           &W   &      1.793 &      0.721 &       69.7 &       0.726 &       0.631 &          19 &     34 &  0.44 \\
1328$+$307 &K   &      5.438 &      3.117 &       83.2 &       0.819 &       0.401 &          24 &     17 &  0.63 \\
           &Q   &      2.682 &      1.494 &       86.0 &       0.067 &       0.061 &           7 &      9 &  0.67 \\
           &W   &      1.473 &      0.705 &  $-$70.7 &       0.084 &       0.081 &           6 &     14 &  0.47 \\
           &D   &      0.966 &      0.507 &  $-$60.5 &       0.189 &       0.190 &          14 &     13 &  0.45 \\
1343$+$451 &K   &      5.258 &      3.243 &       86.2 &       0.820 &       0.819 &           8 &    100 &  0.68 \\
           &Q   &      2.626 &      1.559 &       89.7 &       0.585 &       0.551 &          12 &     46 &  0.57 \\
           &W   &      1.313 &      0.766 &  $-$83.6 &       0.333 &       0.262 &          21 &     12 &  0.48 \\
           &D   &      0.853 &      0.544 &  $-$76.3 &       0.274 &       0.267 &          16 &     17 &  0.60 \\
1611$+$343 &K   &      6.967 &      3.088 &       66.1 &       4.018 &       3.331 &          19 &    172 &  0.67 \\
           &Q   &      3.524 &      1.486 &       67.0 &       2.244 &       2.106 &          34 &     61 &  0.63 \\
           &W   &      1.792 &      0.727 &       68.2 &       1.315 &       1.191 &          38 &     32 &  0.55 \\
           &D   &      1.684 &      0.515 &       65.6 &       0.309 &       0.324 &          70 &      5 &  0.47 \\
1633$+$382 &K   &      6.263 &      3.389 &       65.5 &       5.233 &       5.170 &          34 &    151 &  0.70 \\
           &Q   &      3.303 &      1.644 &       63.6 &       5.500 &       5.502 &          61 &     91 &  0.72 \\
           &W   &      1.705 &      0.818 &       62.5 &       4.837 &       4.476 &          35 &    127 &  0.62 \\
           &D   &      1.023 &      0.538 &       73.8 &       3.056 &       3.002 &          54 &     55 &  0.62 \\
1641$+$399 &K   &      5.053 &      3.521 &  $-$81.7 &       4.866 &       4.200 &          40 &    104 &  0.63 \\
           &Q   &      2.524 &      1.697 &  $-$85.5 &       3.695 &       3.231 &          31 &    106 &  0.62 \\
           &W   &      1.251 &      0.832 &  $-$81.2 &       2.390 &       1.875 &          22 &     85 &  0.66 \\
           &D   &      0.974 &      0.568 &       80.3 &       1.257 &       1.253 &          44 &     29 &  0.59 \\
1749$+$096 &K   &      5.135 &      3.567 &  $-$83.9 &       2.933 &       2.829 &          12 &    230 &  0.62 \\
           &Q   &      2.621 &      1.717 &  $-$86.6 &       2.560 &       2.390 &          13 &    181 &  0.67 \\
           &W   &      1.350 &      0.821 &  $-$80.5 &       1.891 &       1.502 &          26 &     58 &  0.77 \\
           &D   &      0.885 &      0.568 &  $-$78.0 &       1.174 &       1.124 &          41 &     28 &  0.65 \\
1921$-$293 &K   &      6.856 &      5.243 &  $-$22.3 &       7.553 &       7.452 &          77 &     97 &  0.57 \\
           &Q   &      3.466 &      2.604 &  $-$17.4 &       5.092 &       4.822 &          27 &    178 &  0.62 \\
           &W   &      1.841 &      1.252 &  $-$18.8 &       3.012 &       2.246 &          87 &     26 &  0.61 \\
           &D   &      1.320 &      0.798 &  $-$23.4 &       1.549 &       1.542 &          81 &     19 &  0.49 \\
2200$+$420 &K   &      5.575 &      3.196 &       87.6 &       5.239 &       4.573 &          20 &    223 &  0.70 \\
           &Q   &      2.799 &      1.588 &       81.0 &       4.674 &       4.271 &          28 &    152 &  0.64 \\
           &W   &      1.423 &      0.776 &       77.9 &       3.458 &       3.497 &          94 &     37 &  0.70 \\
           &D   &      0.900 &      0.546 &  $-$84.4 &       2.253 &       2.190 &         101 &     22 &  0.61 \\
2223$-$052 &K   &      5.495 &      3.886 &       78.0 &       2.312 &       2.242 &          10 &    222 &  0.56 \\
           &Q   &      2.827 &      1.878 &       81.0 &       1.452 &       1.361 &           6 &    213 &  0.56 \\
           &W   &      1.446 &      0.921 &       82.0 &       0.629 &       0.604 &          16 &     39 &  0.65 \\
           &D   &      0.909 &      0.650 &  $-$70.3 &       0.267 &       0.253 &          22 &     11 &  0.52 \\
2230$+$114 &K   &      5.268 &      3.409 &  $-$88.0 &       2.313 &       2.088 &          10 &    208 &  0.59 \\
           &Q   &      2.689 &      1.642 &       87.6 &       2.138 &       2.023 &          15 &    138 &  0.70 \\
           &W   &      1.354 &      0.792 &       82.1 &       1.506 &       1.519 &          30 &     51 &  0.61 \\
           &D   &      0.876 &      0.564 &  $-$72.3 &       1.568 &       1.089 &          18 &     59 &  0.59 \\
2251$+$158 &K   &      5.205 &      3.419 &  $-$81.5 &       7.036 &       6.044 &          25 &    239 &  0.54 \\
           &Q   &      2.722 &      1.582 &  $-$84.1 &       9.040 &       8.790 &          42 &    209 &  0.61 \\
           &W   &      1.348 &      0.781 &  $-$86.3 &       8.404 &       8.067 &          64 &    126 &  0.75 \\
           &D   &      0.901 &      0.553 &  $-$82.8 &       6.520 &       5.975 &          67 &     89 &  0.62 \\
\enddata
\tablecomments{ 
Column designation: 1~-~source name; 2~-~observing frequency band: 
K~-~22~GHz band; Q~-~43~GHz band; W~-~86~GHz band; D~-~129~GHz band; 
3-5~-~restoring beam:
3~-~major axis [mas];
4~-~minor axis [mas];
5~-~position angle of the major axis [degree];
6~-~total cleaned KVN flux density [Jy];
7~-~peak flux density [Jy beam$^{-1}$];
8~-~off-source RMS in the image [Jy beam$^{-1}$];
9~-~dynamic range of the image;
10~-~quality of the residual noise in the image.
}
\end{deluxetable}

\begin{deluxetable}{lccccrrrr}
\tabletypesize{\scriptsize}
\tablecaption{Model fitting parameters\label{t3}}
\tablewidth{0pt}
\tablehead{
\colhead{Name} & 
\colhead{Band} & 
\colhead{ID} & 
\colhead{$S_{\rm tot}$} &
\colhead{$S_{\rm peak}$} & 
\colhead{$d$} & 
\colhead{$r$} &
\colhead{$\theta$} &
\colhead{$T_{\rm b}$} \\ 
\colhead{(1)} & 
\colhead{(2)} & 
\colhead{(3)} &
\colhead{(4)} & 
\colhead{(5)} & 
\colhead{(6)} &
\colhead{(7)} &
\colhead{(8)} &
\colhead{(9)}
}
\startdata
  0234+285   &  K   & C   &   2.951$\pm$0.111 &   2.816$\pm$0.077 &  0.876$\pm$0.024    &              ... &             ... &   1.394$\pm$ 0.076 \\
             &      & J2  &   0.084$\pm$0.018 &   0.085$\pm$0.013 &  $<${\it 0.602}     &  3.741$\pm$0.046 &     1.4$\pm$0.7 &  $>${\it  0.084} \\
             &  Q   & C   &   2.290$\pm$0.322 &   2.050$\pm$0.215 &  0.629$\pm$0.066    &              ... &             ... &   2.098$\pm$ 0.440 \\
             &      & J1  &   0.111$\pm$0.095 &   0.123$\pm$0.070 &  $<${\it 1.171}     &  1.820$\pm$0.334 &   -26.1$\pm$10.4 &  $>${\it  0.029} \\
             &  W   & C   &   1.522$\pm$0.125 &   1.218$\pm$0.078 &  0.468$\pm$0.030    &              ... &             ... &   2.519$\pm$ 0.323 \\
             &  D   & C   &   0.839$\pm$0.060 &   0.785$\pm$0.041 &  0.183$\pm$0.010    &              ... &             ... &   9.080$\pm$ 0.949 \\
  0235+164   &  K   & C   &   0.884$\pm$0.027 &   0.857$\pm$0.019 &  0.744$\pm$0.016    &              ... &             ... &   0.509$\pm$ 0.022 \\
             &  Q   & C   &   0.846$\pm$0.028 &   0.756$\pm$0.018 &  0.698$\pm$0.017    &              ... &             ... &   0.553$\pm$ 0.027 \\
             &  W   & C   &   0.496$\pm$0.069 &   0.420$\pm$0.045 &  0.408$\pm$0.043    &              ... &             ... &   0.950$\pm$ 0.202 \\
             &  D   & C   &   0.314$\pm$0.149 &   0.275$\pm$0.098 &  0.361$\pm$0.128    &              ... &             ... &   0.768$\pm$ 0.551 \\
  0316+413   &  K   & C   &  28.231$\pm$1.670 &  22.837$\pm$1.050 &  1.770$\pm$0.081    &              ... &             ... &   1.507$\pm$ 0.139 \\
             &      & J1  &   1.481$\pm$0.175 &   1.435$\pm$0.122 &  0.713$\pm$0.061    &  2.991$\pm$0.030 &     1.7$\pm$0.6 &   0.487$\pm$ 0.083 \\
             &      & J2  &   0.408$\pm$0.300 &   0.458$\pm$0.224 &  $<${\it 2.004}     &  6.403$\pm$0.490 &  -129.2$\pm$4.4 &  $>${\it  0.017} \\
             &  Q   & C   &   7.317$\pm$1.670 &   5.894$\pm$1.047 &  0.855$\pm$0.152    &              ... &             ... &   1.673$\pm$ 0.596 \\
             &      & J3  &  11.060$\pm$2.242 &  11.175$\pm$1.593 &  $<${\it 0.281}     &  1.985$\pm$0.020 &  -178.9$\pm$0.6 &  $>${\it 23.345} \\
             &  W   & C   &   6.029$\pm$1.303 &   4.497$\pm$0.779 &  0.517$\pm$0.090    &              ... &             ... &   3.771$\pm$ 1.308 \\
             &      & J3  &   8.149$\pm$1.490 &   3.340$\pm$0.565 &  0.928$\pm$0.157    &  2.413$\pm$0.079 &   179.1$\pm$1.9 &   1.582$\pm$ 0.535 \\
             &  D   & C   &   3.795$\pm$0.329 &   3.246$\pm$0.214 &  0.276$\pm$0.018    &              ... &             ... &   8.329$\pm$ 1.097 \\
             &      & J3  &   2.977$\pm$0.313 &   1.830$\pm$0.164 &  0.487$\pm$0.044    &  2.444$\pm$0.022 &   177.3$\pm$0.5 &   2.099$\pm$ 0.376 \\
             &      & J4  &   0.469$\pm$0.186 &   0.338$\pm$0.109 &  0.441$\pm$0.142    &  1.559$\pm$0.071 &   177.9$\pm$2.6 &   0.403$\pm$ 0.260 \\
             &      & J5  &   0.689$\pm$0.374 &   0.288$\pm$0.144 &  0.700$\pm$0.351    &  0.812$\pm$0.175 &   150.6$\pm$12.2 &   0.235$\pm$ 0.236 \\
  0415+379   &  K   & C   &   3.934$\pm$0.344 &   3.940$\pm$0.243 &  $<${\it 0.255}     &              ... &             ... &  $>${\it 10.466} \\
             &      & J1  &   0.237$\pm$0.119 &   0.250$\pm$0.086 &  $<${\it 1.460}     &  6.446$\pm$0.252 &    64.5$\pm$2.2 &  $>${\it  0.019} \\
             &  Q   & C   &   3.696$\pm$0.394 &   3.708$\pm$0.279 &  $<${\it 0.156}     &              ... &             ... &  $>${\it 26.125} \\
             &  W   & C   &   3.894$\pm$0.090 &   3.712$\pm$0.062 &  0.219$\pm$0.004    &              ... &             ... &  13.994$\pm$ 0.471 \\
             &  D   & C   &   2.749$\pm$0.047 &   2.688$\pm$0.033 &  0.102$\pm$0.001    &              ... &             ... &  45.541$\pm$ 1.124 \\
  0420-014   &  K   & C   &   5.113$\pm$0.064 &   5.058$\pm$0.045 &  0.463$\pm$0.004    &              ... &             ... &   7.508$\pm$ 0.134 \\
             &  Q   & C   &   4.046$\pm$0.132 &   3.777$\pm$0.090 &  0.574$\pm$0.014    &              ... &             ... &   3.866$\pm$ 0.184 \\
             &  W   & C   &   2.402$\pm$0.711 &   2.453$\pm$0.508 &  $<${\it 0.227}     &              ... &             ... &  $>${\it 14.685} \\
             &  D   & C   &   1.643$\pm$0.164 &   1.616$\pm$0.115 &  0.110$\pm$0.008    &              ... &             ... &  42.745$\pm$ 6.083 \\
  0528+134   &  K   & C   &   1.147$\pm$0.083 &   1.149$\pm$0.059 &  $<${\it 0.203}     &              ... &             ... &  $>${\it 14.053} \\
             &      & J1  &   0.179$\pm$0.096 &   0.066$\pm$0.033 &  4.406$\pm$2.224    &  4.543$\pm$1.112 &    37.6$\pm$13.8 &   0.005$\pm$ 0.005 \\
             &  Q   & C   &   0.935$\pm$0.045 &   0.878$\pm$0.031 &  0.520$\pm$0.018    &              ... &             ... &   1.744$\pm$ 0.122 \\
             &  W   & C   &   0.614$\pm$0.054 &   0.519$\pm$0.035 &  0.420$\pm$0.028    &              ... &             ... &   1.756$\pm$ 0.238 \\
             &  D   & C   &   0.333$\pm$0.069 &   0.336$\pm$0.049 &  $<${\it 0.097}     &              ... &             ... &  $>${\it 17.823} \\
  0716+714   &  K   & C   &   2.490$\pm$0.465 &   2.511$\pm$0.330 &  $<${\it 0.616}     &              ... &             ... &  $>${\it  1.213} \\
             &  Q   & C   &   2.733$\pm$0.047 &   2.654$\pm$0.033 &  0.427$\pm$0.005    &              ... &             ... &   2.775$\pm$ 0.069 \\
             &  W   & C   &   2.061$\pm$0.041 &   2.045$\pm$0.029 &  0.109$\pm$0.002    &              ... &             ... &  32.119$\pm$ 0.902 \\
             &  D   & C   &   1.436$\pm$0.034 &   1.425$\pm$0.024 &  0.073$\pm$0.001    &              ... &             ... &  49.894$\pm$ 1.679 \\
  0727-115   &  K   & C   &   3.848$\pm$0.126 &   3.724$\pm$0.088 &  0.866$\pm$0.020    &              ... &             ... &   2.184$\pm$ 0.103 \\
             &      & J1  &   0.101$\pm$0.048 &   0.108$\pm$0.035 &  $<${\it 1.587}     &  3.822$\pm$0.258 &   -82.9$\pm$3.9 &  $>${\it  0.017} \\
             &  Q   & C   &   2.367$\pm$0.044 &   2.292$\pm$0.031 &  0.428$\pm$0.006    &              ... &             ... &   5.500$\pm$ 0.148 \\
             &  W   & C   &   1.146$\pm$0.198 &   1.155$\pm$0.141 &  $<${\it 0.143}     &              ... &             ... &  $>${\it 23.705} \\
             &  D   & C   &   0.786$\pm$0.025 &   0.766$\pm$0.018 &  0.129$\pm$0.003    &              ... &             ... &  20.106$\pm$ 0.931 \\
  0735+178   &  K   & C   &   1.000$\pm$0.090 &   1.002$\pm$0.063 &  $<${\it 0.274}     &              ... &             ... &  $>${\it  3.167} \\
             &      & J1  &   0.104$\pm$0.033 &   0.052$\pm$0.015 &  3.588$\pm$1.005    &  4.290$\pm$0.502 &    71.7$\pm$6.7 &   0.002$\pm$ 0.001 \\
             &  Q   & C   &   0.908$\pm$0.045 &   0.819$\pm$0.030 &  0.659$\pm$0.024    &              ... &             ... &   0.498$\pm$ 0.037 \\
             &  W   & C   &   0.738$\pm$0.079 &   0.551$\pm$0.047 &  0.550$\pm$0.047    &              ... &             ... &   0.581$\pm$ 0.099 \\
             &  D   & C   &   0.426$\pm$0.031 &   0.380$\pm$0.021 &  0.229$\pm$0.012    &              ... &             ... &   1.935$\pm$ 0.211 \\
             &      & J1  &   0.068$\pm$0.017 &   0.068$\pm$0.012 &  $<${\it 0.118}     &  0.645$\pm$0.010 &    29.7$\pm$0.9 &  $>${\it  1.169} \\
  0827+243   &  K   & C   &   0.905$\pm$0.027 &   0.868$\pm$0.019 &  0.912$\pm$0.020    &              ... &             ... &   0.347$\pm$ 0.015 \\
             &  Q   & C   &   0.667$\pm$0.039 &   0.592$\pm$0.026 &  0.727$\pm$0.032    &              ... &             ... &   0.403$\pm$ 0.035 \\
             &  W   & C   &   0.324$\pm$0.028 &   0.267$\pm$0.018 &  0.467$\pm$0.032    &              ... &             ... &   0.474$\pm$ 0.064 \\
             &  D   & C   &   0.287$\pm$0.015 &   0.275$\pm$0.011 &  0.158$\pm$0.006    &              ... &             ... &   3.668$\pm$ 0.286 \\
  0836+710   &  K   & C   &   1.994$\pm$0.141 &   1.633$\pm$0.089 &  2.265$\pm$0.124    &              ... &             ... &   0.205$\pm$ 0.023 \\
             &      & J1  &   0.061$\pm$0.027 &   0.065$\pm$0.020 &  $<${\it 1.470}     &  4.698$\pm$0.227 &  -136.9$\pm$2.8 &  $>${\it  0.015} \\
             &  Q   & C   &   1.065$\pm$0.029 &   1.048$\pm$0.021 &  0.317$\pm$0.006    &              ... &             ... &   5.603$\pm$ 0.221 \\
             &      & J2  &   0.137$\pm$0.017 &   0.137$\pm$0.012 &  0.213$\pm$0.018    &  2.458$\pm$0.009 &  -140.1$\pm$0.2 &   1.596$\pm$ 0.274 \\
             &  W   & C   &   0.656$\pm$0.160 &   0.590$\pm$0.107 &  0.460$\pm$0.083    &              ... &             ... &   1.639$\pm$ 0.596 \\
             &  D   & C   &   0.519$\pm$0.060 &   0.335$\pm$0.032 &  0.566$\pm$0.055    &              ... &             ... &   0.857$\pm$ 0.165 \\
             &      & J3  &   0.053$\pm$0.007 &   0.044$\pm$0.005 &  0.359$\pm$0.038    &  0.954$\pm$0.019 &  -153.8$\pm$1.1 &   0.217$\pm$ 0.046 \\
  0851+202   &  K   & C   &   4.707$\pm$0.435 &   4.714$\pm$0.307 &  $<${\it 0.266}     &              ... &             ... &  $>${\it 14.244} \\
             &      & J1  &   0.252$\pm$0.134 &   0.123$\pm$0.059 &  3.476$\pm$1.659    &  4.026$\pm$0.829 &  -107.2$\pm$11.6 &   0.004$\pm$ 0.004 \\
             &  Q   & C   &   3.838$\pm$0.089 &   3.683$\pm$0.062 &  0.414$\pm$0.007    &              ... &             ... &   4.805$\pm$ 0.162 \\
             &  W   & C   &   2.755$\pm$0.112 &   2.453$\pm$0.074 &  0.345$\pm$0.010    &              ... &             ... &   4.966$\pm$ 0.301 \\
             &  D   & C   &   1.764$\pm$0.106 &   1.653$\pm$0.073 &  0.179$\pm$0.008    &              ... &             ... &  11.813$\pm$ 1.041 \\
  0923+392   &  K   & C   &   0.552$\pm$0.123 &   6.518$\pm$0.123 &  3.812$\pm$0.072    &              ... &             ... &   0.011$\pm$ 0.001 \\
             &      & J1  &   7.994$\pm$18.767 &   0.262$\pm$0.615 &  $<${\it 0.326}     &  5.683$\pm$0.383 &   100.3$\pm$3.9 &  $>${\it 20.917} \\
             &  Q   & C   &   0.743$\pm$0.127 &   0.237$\pm$0.039 &  1.147$\pm$0.187    &              ... &             ... &   0.157$\pm$ 0.051 \\
             &      & J1  &   5.222$\pm$2.652 &   0.555$\pm$0.280 &  0.607$\pm$0.307    &  3.174$\pm$0.153 &   100.6$\pm$2.8 &   3.947$\pm$ 3.986 \\
             &  W   & C   &   0.509$\pm$0.056 &   0.501$\pm$0.039 &  0.755$\pm$0.059    &              ... &             ... &   0.249$\pm$ 0.039 \\
             &      & J1  &   3.309$\pm$2.701 &   0.301$\pm$0.245 &  0.629$\pm$0.511    &  2.779$\pm$0.256 &    98.3$\pm$5.3 &   2.329$\pm$ 3.787 \\
             &  D   & C   &   1.899$\pm$0.306 &   1.141$\pm$0.158 &  0.555$\pm$0.077    &              ... &             ... &   1.717$\pm$ 0.475 \\
  1055+018   &  K   & C   &   2.849$\pm$0.262 &   2.856$\pm$0.185 &  $<${\it 0.269}     &              ... &             ... &  $>${\it 12.239} \\
             &  Q   & C   &   2.438$\pm$0.071 &   2.348$\pm$0.049 &  0.438$\pm$0.009    &              ... &             ... &   3.942$\pm$ 0.166 \\
             &  W   & C   &   1.717$\pm$0.064 &   1.475$\pm$0.041 &  0.426$\pm$0.012    &              ... &             ... &   2.935$\pm$ 0.165 \\
             &  D   & C   &   1.068$\pm$0.030 &   1.047$\pm$0.021 &  0.103$\pm$0.002    &              ... &             ... &  31.226$\pm$ 1.233 \\
  1101+384   &  K   & C   &   0.442$\pm$0.019 &   0.435$\pm$0.013 &  0.530$\pm$0.016    &              ... &             ... &   0.266$\pm$ 0.016 \\
             &  Q   & C   &   0.347$\pm$0.018 &   0.311$\pm$0.012 &  0.661$\pm$0.025    &              ... &             ... &   0.134$\pm$ 0.010 \\
             &  W   & C   &   0.222$\pm$0.052 &   0.187$\pm$0.034 &  0.438$\pm$0.079    &              ... &             ... &   0.196$\pm$ 0.071 \\
             &  D   & C   &   0.365$\pm$0.018 &   0.348$\pm$0.012 &  0.151$\pm$0.005    &              ... &             ... &   2.711$\pm$ 0.188 \\
  1127-145   &  K   & C   &   2.074$\pm$0.144 &   2.068$\pm$0.102 &  $<${\it 0.273}     &              ... &             ... &  $>${\it 10.012} \\
             &  Q   & C   &   1.837$\pm$0.213 &   1.794$\pm$0.149 &  0.464$\pm$0.039    &              ... &             ... &   3.062$\pm$ 0.509 \\
             &  W   & C   &   1.391$\pm$0.285 &   1.406$\pm$0.203 &  $<${\it 0.197}     &              ... &             ... &  $>${\it 12.913} \\
             &  D   & C   &   0.747$\pm$0.087 &   0.749$\pm$0.061 &  $<${\it 0.074}     &              ... &             ... &  $>${\it 48.720} \\
  1156+295   &  K   & C   &   0.866$\pm$0.027 &   0.853$\pm$0.019 &  0.496$\pm$0.011    &              ... &             ... &   0.998$\pm$ 0.044 \\
             &  Q   & C   &   0.892$\pm$0.044 &   0.809$\pm$0.030 &  0.624$\pm$0.023    &              ... &             ... &   0.649$\pm$ 0.048 \\
             &  W   & C   &   0.790$\pm$0.064 &   0.635$\pm$0.040 &  0.462$\pm$0.029    &              ... &             ... &   1.049$\pm$ 0.132 \\
             &  D   & C   &   0.490$\pm$0.032 &   0.481$\pm$0.022 &  0.095$\pm$0.004    &              ... &             ... &  15.387$\pm$ 1.418 \\
  1222+216   &  K   & C   &   2.351$\pm$0.044 &   2.287$\pm$0.031 &  0.713$\pm$0.010    &              ... &             ... &   1.090$\pm$ 0.029 \\
             &  Q   & C   &   1.734$\pm$0.090 &   1.495$\pm$0.059 &  0.792$\pm$0.031    &              ... &             ... &   0.651$\pm$ 0.051 \\
             &      & J1  &   0.175$\pm$0.019 &   0.171$\pm$0.013 &  0.356$\pm$0.028    &  1.603$\pm$0.014 &    13.6$\pm$0.5 &   0.325$\pm$ 0.051 \\
             &  W   & C   &   0.872$\pm$0.109 &   0.776$\pm$0.073 &  0.336$\pm$0.031    &              ... &             ... &   1.820$\pm$ 0.341 \\
             &      & J1  &   0.253$\pm$0.072 &   0.258$\pm$0.051 &  $<${\it 0.195}     &  0.981$\pm$0.019 &     4.2$\pm$1.1 &  $>${\it  1.575} \\
             &  D   & C   &   0.403$\pm$0.136 &   0.426$\pm$0.099 &  $<${\it 0.187}     &              ... &             ... &  $>${\it  2.708} \\
             &      & J1  &   0.123$\pm$0.057 &   0.123$\pm$0.040 &  $<${\it 0.247}     &  0.925$\pm$0.040 &   -14.1$\pm$2.5 &  $>${\it  0.473} \\
  1226+023   &  K   & C   &   6.945$\pm$0.810 &   6.391$\pm$0.548 &  1.307$\pm$0.112    &              ... &             ... &   0.774$\pm$ 0.133 \\
             &      & J3  &   1.511$\pm$0.375 &   1.538$\pm$0.268 &  $<${\it 0.745}     &  4.120$\pm$0.065 &  -136.2$\pm$0.9 &  $>${\it  0.519} \\
             &  Q   & C   &   4.749$\pm$0.062 &   4.688$\pm$0.044 &  0.259$\pm$0.002    &              ... &             ... &  13.472$\pm$ 0.251 \\
             &      & J2  &   0.823$\pm$0.041 &   0.800$\pm$0.029 &  0.377$\pm$0.014    &  1.962$\pm$0.007 &  -140.2$\pm$0.2 &   1.102$\pm$ 0.079 \\
             &      & J3  &   0.958$\pm$0.054 &   0.675$\pm$0.031 &  1.386$\pm$0.064    &  4.344$\pm$0.032 &  -143.5$\pm$0.4 &   0.095$\pm$ 0.009 \\
             &      & J4  &   1.373$\pm$0.212 &   0.350$\pm$0.052 &  3.064$\pm$0.459    &  7.402$\pm$0.230 &  -141.9$\pm$1.8 &   0.028$\pm$ 0.008 \\
             &      & J5  &   0.720$\pm$0.161 &   0.291$\pm$0.060 &  2.353$\pm$0.487    &  10.642$\pm$0.243 &  -141.6$\pm$1.3 &   0.025$\pm$ 0.010 \\
             &  W   & C   &   2.987$\pm$0.248 &   2.144$\pm$0.145 &  0.641$\pm$0.043    &              ... &             ... &   1.383$\pm$ 0.187 \\
             &      & J1  &   1.293$\pm$0.246 &   0.852$\pm$0.135 &  0.709$\pm$0.113    &  1.043$\pm$0.056 &  -138.2$\pm$3.1 &   0.489$\pm$ 0.156 \\
             &      & J2  &   0.184$\pm$0.096 &   0.198$\pm$0.071 &  $<${\it 0.398}     &  1.931$\pm$0.071 &  -143.1$\pm$2.1 &  $>${\it  0.221} \\
             &  D   & C   &   1.829$\pm$0.225 &   1.605$\pm$0.149 &  0.279$\pm$0.026    &              ... &             ... &   4.471$\pm$ 0.829 \\
             &      & J1  &   0.936$\pm$0.403 &   0.427$\pm$0.167 &  0.704$\pm$0.276    &  0.903$\pm$0.138 &  -143.1$\pm$8.7 &   0.359$\pm$ 0.282 \\
  1228+126   &  K   & C   &   1.837$\pm$0.210 &   1.844$\pm$0.149 &  $<${\it 0.321}     &              ... &             ... &  $>${\it  2.933} \\
             &      & J1  &   0.199$\pm$0.013 &   0.198$\pm$0.009 &  0.408$\pm$0.018    &  5.115$\pm$0.009 &   -69.7$\pm$0.1 &   0.197$\pm$ 0.018 \\
             &  Q   & C   &   1.286$\pm$0.197 &   1.295$\pm$0.140 &  $<${\it 0.213}     &              ... &             ... &  $>${\it  4.690} \\
             &      & J1  &   0.098$\pm$0.034 &   0.090$\pm$0.023 &  $<${\it 0.449}     &  3.023$\pm$0.057 &   -70.0$\pm$1.1 &  $>${\it  0.080} \\
             &  W   & C   &   1.068$\pm$0.049 &   0.947$\pm$0.033 &  0.362$\pm$0.012    &              ... &             ... &   1.345$\pm$ 0.093 \\
             &  D   & C   &   0.657$\pm$0.073 &   0.659$\pm$0.051 &  $<${\it 0.054}     &              ... &             ... &  $>${\it 37.271} \\
  1253-055   &  K   & C   &  30.173$\pm$1.175 &  26.605$\pm$0.777 &  1.882$\pm$0.055    &              ... &             ... &   2.150$\pm$ 0.126 \\
             &  Q   & C   &  20.803$\pm$0.811 &  18.545$\pm$0.540 &  0.857$\pm$0.025    &              ... &             ... &   7.148$\pm$ 0.416 \\
             &      & J1  &   1.888$\pm$0.356 &   1.720$\pm$0.240 &  0.843$\pm$0.118    &  2.106$\pm$0.059 &  -137.2$\pm$1.6 &   0.670$\pm$ 0.187 \\
             &  W   & C   &   7.396$\pm$1.100 &   6.780$\pm$0.743 &  0.300$\pm$0.033    &  0.287$\pm$0.016 &    28.3$\pm$3.3 &  20.738$\pm$ 4.553 \\
             &      & J1  &   5.256$\pm$1.195 &   5.320$\pm$0.850 &  $<${\it 0.183}     &  0.454$\pm$0.015 &  -152.8$\pm$1.8 &  $>${\it 39.733} \\
             &  D   & C   &   7.699$\pm$1.481 &   4.123$\pm$0.699 &  0.591$\pm$0.100    &              ... &             ... &   5.562$\pm$ 1.887 \\
             &      & J1  &   2.372$\pm$0.688 &   2.426$\pm$0.492 &  $<${\it 0.153}     &  0.726$\pm$0.016 &  -153.7$\pm$1.2 &  $>${\it 25.569} \\
  1308+326   &  K   & C   &   1.436$\pm$0.405 &   1.464$\pm$0.289 &  $<${\it 0.851}     &              ... &             ... &  $>${\it  0.650} \\
             &  Q   & C   &   1.105$\pm$0.030 &   1.055$\pm$0.021 &  0.434$\pm$0.009    &              ... &             ... &   1.925$\pm$ 0.077 \\
             &  W   & C   &   0.740$\pm$0.113 &   0.615$\pm$0.072 &  0.442$\pm$0.052    &              ... &             ... &   1.243$\pm$ 0.292 \\
  1328+307   &  K   & C   &   0.602$\pm$0.194 &   0.375$\pm$0.103 &  2.917$\pm$0.798    &              ... &             ... &   0.021$\pm$ 0.012 \\
             &      & J1  &   0.453$\pm$0.070 &   0.283$\pm$0.037 &  2.875$\pm$0.376    &  5.248$\pm$0.188 &    43.2$\pm$2.1 &   0.017$\pm$ 0.004 \\
             &  Q   & C   &   0.083$\pm$0.015 &   0.061$\pm$0.009 &  1.106$\pm$0.166    &              ... &             ... &   0.021$\pm$ 0.006 \\
             &  W   & C   &   0.102$\pm$0.010 &   0.081$\pm$0.006 &  0.472$\pm$0.037    &              ... &             ... &   0.139$\pm$ 0.022 \\
             &  D   & C   &   0.195$\pm$0.020 &   0.190$\pm$0.014 &  0.122$\pm$0.009    &              ... &             ... &   3.973$\pm$ 0.586 \\
  1343+451   &  K   & C   &   0.826$\pm$0.026 &   0.819$\pm$0.018 &  0.382$\pm$0.008    &              ... &             ... &   3.287$\pm$ 0.146 \\
             &  Q   & C   &   0.607$\pm$0.037 &   0.549$\pm$0.025 &  0.649$\pm$0.029    &              ... &             ... &   0.837$\pm$ 0.075 \\
             &  W   & C   &   0.326$\pm$0.060 &   0.250$\pm$0.036 &  0.499$\pm$0.073    &              ... &             ... &   0.760$\pm$ 0.221 \\
             &  D   & C   &   0.279$\pm$0.024 &   0.267$\pm$0.017 &  0.148$\pm$0.009    &              ... &             ... &   7.395$\pm$ 0.927 \\
  1611+343   &  K   & C   &   3.229$\pm$0.386 &   3.242$\pm$0.273 &  $<${\it 0.368}     &              ... &             ... &  $>${\it  9.380} \\
             &      & J1  &   0.723$\pm$0.129 &   0.582$\pm$0.081 &  1.940$\pm$0.271    &  4.360$\pm$0.135 &   163.9$\pm$1.8 &   0.076$\pm$ 0.021 \\
             &      & J2  &   0.258$\pm$0.102 &   0.173$\pm$0.057 &  3.041$\pm$1.001    &  3.443$\pm$0.500 &  -132.2$\pm$8.3 &   0.011$\pm$ 0.007 \\
             &  Q   & C   &   2.317$\pm$0.144 &   2.085$\pm$0.096 &  0.678$\pm$0.031    &              ... &             ... &   1.987$\pm$ 0.184 \\
             &  W   & C   &   1.508$\pm$0.126 &   1.186$\pm$0.078 &  0.509$\pm$0.033    &              ... &             ... &   2.295$\pm$ 0.301 \\
             &  D   & C   &   0.327$\pm$0.036 &   0.324$\pm$0.026 &  0.097$\pm$0.008    &              ... &             ... &  13.703$\pm$ 2.171 \\
  1633+382   &  K   & C   &   5.266$\pm$0.104 &   5.172$\pm$0.073 &  0.608$\pm$0.009    &              ... &             ... &   6.584$\pm$ 0.185 \\
             &  Q   & C   &   5.856$\pm$0.112 &   5.503$\pm$0.077 &  0.559$\pm$0.008    &              ... &             ... &   8.661$\pm$ 0.241 \\
             &  W   & C   &   5.104$\pm$0.188 &   4.457$\pm$0.124 &  0.416$\pm$0.012    &              ... &             ... &  13.630$\pm$ 0.757 \\
             &  D   & C   &   3.247$\pm$0.145 &   3.004$\pm$0.099 &  0.202$\pm$0.007    &              ... &             ... &  36.776$\pm$ 2.418 \\
  1641+399   &  K   & C   &   4.406$\pm$0.236 &   4.184$\pm$0.163 &  0.998$\pm$0.039    &              ... &             ... &   1.158$\pm$ 0.090 \\
             &      & J4  &   0.960$\pm$0.333 &   0.526$\pm$0.160 &  3.537$\pm$1.077    &  5.955$\pm$0.538 &   -84.5$\pm$5.2 &   0.020$\pm$ 0.012 \\
             &  Q   & C   &   3.432$\pm$0.210 &   3.225$\pm$0.144 &  0.534$\pm$0.024    &              ... &             ... &   3.150$\pm$ 0.281 \\
             &      & J3  &   0.372$\pm$0.102 &   0.308$\pm$0.065 &  0.940$\pm$0.198    &  2.667$\pm$0.099 &   -90.7$\pm$2.1 &   0.110$\pm$ 0.047 \\
             &  W   & C   &   1.930$\pm$0.141 &   1.879$\pm$0.098 &  0.162$\pm$0.008    &              ... &             ... &  19.247$\pm$ 2.016 \\
             &      & J1  &   0.118$\pm$0.038 &   0.123$\pm$0.027 &  $<${\it 0.219}     &  0.787$\pm$0.024 &  -128.9$\pm$1.8 &  $>${\it  0.644} \\
             &      & J2  &   0.363$\pm$0.102 &   0.370$\pm$0.073 &  $<${\it 0.190}     &  1.340$\pm$0.019 &   -92.2$\pm$0.8 &  $>${\it  2.630} \\
             &      & J3  &   0.159$\pm$0.032 &   0.148$\pm$0.022 &  0.316$\pm$0.047    &  2.169$\pm$0.024 &   -76.7$\pm$0.6 &   0.417$\pm$ 0.124 \\
             &  D   & C   &   1.215$\pm$0.392 &   1.254$\pm$0.281 &  $<${\it 0.160}     &              ... &             ... &  $>${\it 12.434} \\
  1749+096   &  K   & C   &   2.799$\pm$0.352 &   2.808$\pm$0.249 &  $<${\it 0.357}     &              ... &             ... &  $>${\it  4.766} \\
             &      & J1  &   0.145$\pm$0.082 &   0.091$\pm$0.043 &  3.503$\pm$1.671    &  3.845$\pm$0.835 &    48.9$\pm$12.3 &   0.003$\pm$ 0.002 \\
             &  Q   & C   &   2.718$\pm$0.158 &   2.382$\pm$0.104 &  0.795$\pm$0.035    &              ... &             ... &   0.934$\pm$ 0.082 \\
             &  W   & C   &   1.700$\pm$0.523 &   1.465$\pm$0.341 &  0.470$\pm$0.110    &              ... &             ... &   1.671$\pm$ 0.782 \\
             &  D   & C   &   1.273$\pm$0.318 &   1.125$\pm$0.211 &  0.284$\pm$0.053    &              ... &             ... &   3.428$\pm$ 1.288 \\
  1921-293   &  K   & C   &   8.091$\pm$0.297 &   7.456$\pm$0.201 &  1.808$\pm$0.049    &              ... &             ... &   0.550$\pm$ 0.030 \\
             &  Q   & C   &   5.223$\pm$0.354 &   4.816$\pm$0.240 &  0.888$\pm$0.044    &              ... &             ... &   1.472$\pm$ 0.147 \\
             &      & J2  &   0.271$\pm$0.103 &   0.281$\pm$0.074 &  $<${\it 0.761}     &  3.089$\pm$0.101 &    24.2$\pm$1.9 &  $>${\it  0.104} \\
             &  W   & C   &   2.004$\pm$0.553 &   2.043$\pm$0.395 &  $<${\it 0.279}     &              ... &             ... &  $>${\it  5.739} \\
             &      & J1  &   1.170$\pm$0.351 &   0.609$\pm$0.162 &  1.280$\pm$0.340    &  1.428$\pm$0.170 &     4.2$\pm$6.8 &   0.159$\pm$ 0.084 \\
             &  D   & C   &   1.697$\pm$0.087 &   1.543$\pm$0.058 &  0.320$\pm$0.012    &              ... &             ... &   3.683$\pm$ 0.278 \\
  2200+420   &  K   & C   &   4.838$\pm$0.322 &   4.544$\pm$0.220 &  1.055$\pm$0.051    &              ... &             ... &   0.763$\pm$ 0.074 \\
             &      & J2  &   0.605$\pm$0.087 &   0.499$\pm$0.055 &  1.859$\pm$0.206    &  4.017$\pm$0.103 &  -163.8$\pm$1.5 &   0.031$\pm$ 0.007 \\
             &  Q   & C   &   4.351$\pm$0.164 &   4.231$\pm$0.114 &  0.350$\pm$0.009    &              ... &             ... &   6.236$\pm$ 0.337 \\
             &      & J1  &   0.499$\pm$0.112 &   0.283$\pm$0.055 &  1.623$\pm$0.316    &  2.038$\pm$0.158 &  -161.0$\pm$4.4 &   0.033$\pm$ 0.013 \\
             &  W   & C   &   3.925$\pm$0.094 &   3.498$\pm$0.063 &  0.352$\pm$0.006    &              ... &             ... &   5.561$\pm$ 0.199 \\
             &  D   & C   &   2.326$\pm$0.122 &   2.191$\pm$0.084 &  0.174$\pm$0.007    &              ... &             ... &  13.488$\pm$ 1.029 \\
  2223-052   &  K   & C   &   2.240$\pm$0.164 &   2.244$\pm$0.116 &  $<${\it 0.225}     &              ... &             ... &  $>${\it 17.503} \\
             &      & J2  &   0.084$\pm$0.029 &   0.087$\pm$0.021 &  $<${\it 1.075}     &  6.026$\pm$0.130 &    94.0$\pm$1.2 &  $>${\it  0.029} \\
             &  Q   & C   &   1.427$\pm$0.134 &   1.360$\pm$0.093 &  0.503$\pm$0.034    &              ... &             ... &   2.228$\pm$ 0.304 \\
             &      & J1  &   0.136$\pm$0.042 &   0.094$\pm$0.024 &  1.521$\pm$0.383    &  2.632$\pm$0.191 &   107.0$\pm$4.2 &   0.023$\pm$ 0.012 \\
             &  W   & C   &   0.691$\pm$0.057 &   0.600$\pm$0.037 &  0.443$\pm$0.027    &              ... &             ... &   1.391$\pm$ 0.173 \\
             &  D   & C   &   0.298$\pm$0.046 &   0.253$\pm$0.030 &  0.326$\pm$0.039    &              ... &             ... &   1.107$\pm$ 0.263 \\
  2230+114   &  K   & C   &   2.227$\pm$0.097 &   2.119$\pm$0.067 &  0.968$\pm$0.031    &              ... &             ... &   0.795$\pm$ 0.050 \\
             &      & J1  &   0.299$\pm$0.056 &   0.189$\pm$0.030 &  3.013$\pm$0.480    &  5.372$\pm$0.240 &   167.2$\pm$2.6 &   0.011$\pm$ 0.004 \\
             &  Q   & C   &   2.235$\pm$0.213 &   1.995$\pm$0.142 &  0.719$\pm$0.051    &              ... &             ... &   1.447$\pm$ 0.206 \\
             &  W   & C   &   1.618$\pm$0.083 &   1.519$\pm$0.057 &  0.260$\pm$0.010    &              ... &             ... &   8.010$\pm$ 0.600 \\
             &  D   & C   &   1.308$\pm$0.171 &   1.078$\pm$0.109 &  0.325$\pm$0.033    &              ... &             ... &   4.144$\pm$ 0.838 \\
             &      & J1  &   0.199$\pm$0.020 &   0.190$\pm$0.014 &  0.144$\pm$0.011    &  0.761$\pm$0.005 &    73.8$\pm$0.4 &   3.212$\pm$ 0.478 \\
             &      & J2  &   0.165$\pm$0.064 &   0.170$\pm$0.046 &  $<${\it 0.181}     &  1.046$\pm$0.025 &   115.8$\pm$1.3 &  $>${\it  1.677} \\
  2251+158   &  K   & C   &   5.924$\pm$0.499 &   5.936$\pm$0.353 &  $<${\it 0.236}     &              ... &             ... &  $>${\it 32.440} \\
             &      & J2  &   0.582$\pm$0.100 &   0.485$\pm$0.064 &  1.842$\pm$0.244    &  3.856$\pm$0.122 &   -92.3$\pm$1.8 &   0.052$\pm$ 0.014 \\
             &      & J3  &   0.368$\pm$0.161 &   0.388$\pm$0.117 &  $<${\it 1.234}     &  7.690$\pm$0.186 &   -69.3$\pm$1.4 &  $>${\it  0.074} \\
             &      & J4  &   0.249$\pm$0.053 &   0.224$\pm$0.036 &  1.458$\pm$0.232    &  11.371$\pm$0.116 &   -58.6$\pm$0.6 &   0.036$\pm$ 0.011 \\
             &  Q   & C   &   8.868$\pm$0.190 &   8.721$\pm$0.133 &  0.268$\pm$0.004    &              ... &             ... &  37.709$\pm$ 1.152 \\
             &      & J1  &   0.555$\pm$0.184 &   0.217$\pm$0.067 &  2.128$\pm$0.658    &  2.062$\pm$0.329 &  -120.1$\pm$9.1 &   0.037$\pm$ 0.023 \\
             &  W   & C   &   8.954$\pm$0.424 &   8.073$\pm$0.284 &  0.331$\pm$0.012    &              ... &             ... &  24.961$\pm$ 1.756 \\
             &  D   & C   &   6.124$\pm$0.390 &   5.933$\pm$0.271 &  0.127$\pm$0.006    &              ... &             ... &  115.964$\pm$10.610 \\
\enddata
\tablecomments{ 
Column designation:
1~-~source name;
2~-~observing frequency band: 
K~-~22~GHz band; Q~-~43~GHz band; W~-~86~GHz band; D~-~129~GHz band; 
3~-~component identification: C~-~core component; J~-~jet component;
4~-~model flux density of the component [Jy];
5~-~peak brightness of individual component measured in the image [Jy beam$^{-1}$];
6~-~size [mas], italic numbers indicate upper limits;
7~-~radius [mas];
8~-~position angle [deg];
9~-~measured brightness temperature [$10\times9$~K], italic numbers indicate lower limits.
}
\end{deluxetable}

\end{document}